\documentclass[10pt,twocolumn,preprintnumbers,amsmath,amssymb,nofootinbib,superscriptaddress]{revtex4-1}

\usepackage[english]{babel}
\usepackage{float} 
\usepackage{booktabs}
\usepackage{orcidlink}
\usepackage[letterpaper,top=1cm,bottom=1.5cm,left=1.5cm,right=1.5cm,marginparwidth=1.75cm]{geometry}

\usepackage{hyperref}
\usepackage{amsmath, amssymb, enumitem}
\usepackage{graphicx}
\usepackage{color}
\usepackage[]{mdframed}
\usepackage{comment}
\usepackage[normalem]{ulem}


\newcommand{\bea}{\begin{equation}}
\newcommand{\eea}{\end{equation}}

\begin{document}
\title{Tests of scalar polarizations with multi-messenger events}

\author{Sk Md Adil Imam\orcidlink{0000-0003-3308-2615}}
% \href{https://orcid.org/0000-0003-3308-2615}{\includegraphics[width=0.8em]{Orcid-ID.png}}}
 \email{adil.imam@unab.cl}
  \affiliation{Institute of Astrophysics, Department of Physics and Astronomy, Universidad Andrés Bello, Santiago, Chile}
   \author{Macarena Lagos\orcidlink{0000-0003-0234-9970}}
 \email{macarena.lagos.u@unab.cl}
  \affiliation{Institute of Astrophysics, Department of Physics and Astronomy, Universidad Andrés Bello, Santiago, Chile}

\begin{abstract}
Gravitational wave (GW) observations provide a unique opportunity to test Einstein's General Relativity (GR) in the strong-field regime. While GR predicts only two tensor polarization modes, generic metric theories allow up to six independent modes. We perform a parameterized test of GR using the parameterized post-Einsteinian (PPE) framework applied to GW170817, incorporating for the first time the polarization angle constraints from the gamma-ray burst afterglow alongside other electromagnetic (EM) counterpart information. We extend the GR waveform by adding a scalar breathing mode and modifications to the tensor modes, introducing three non-GR parameters. We perform Bayesian inference for both quadrupole  $\ell = |m|= 2$  and dipole  $\ell = |m|= 1$  angular harmonics, with two frequency evolution models. For $\ell = |m| = 2$, we find that within the extended PPE framework the scalar amplitude deviates from zero at the $2$--$3\sigma$ level, leading to a modest preference for modified gravity. However, due to penalization of extra parameters, Bayesian model comparison still favors pure GR over the extended PPE waveform model, with a log Bayes factor of $\Delta \log \mathcal{Z} = 4.67 \pm 0.32$. The EM constraint on the polarization angle places very tight bounds on non-GR parameters; for instance, in the case $\ell = |m| = 2$, the bound on the scalar (tensor) amplitude modification parameter improves by roughly $60\%$ $(30\%)$, highlighting the impact that long-term follow up of GW events can have on tests of gravity.  
\end{abstract}

\keywords{gravitational waves, modified gravity}

\maketitle
%\tableofcontents

%%%%%%%%%%%%%%%%%%%%%%%%%%%%%%%%%%
%INTRODUCTION
%%%%%%%%%%%%%%%%%%%%%%%%%%%%%%%%%%
\section{Introduction}
While the theory of General Relativity (GR) has been exquisitely confirmed around the Solar System by constraining deviations to less than the $10^{-4}$ level \cite{Will:2014kxa}, its predictive power reaches its limits at high energies, where GR is known to be incomplete \cite{Donoghue:1994dn}. 
Gravitational waves (GWs) are a key observable that provides direct information on the high-energy regime of gravity, since the signals detectable to date come from extreme astrophysical compact objects with strong gravitational fields~\cite{Yunes_2013,Yunes_2016,Ezquiaga_2018}. GW data from the current detector network (LIGO~\cite{Ligo2015}, VIRGO~\cite{Acernese_2014}, KAGRA~\cite{Somiya_2012}) have already been used to perform various tests of gravity~\cite{Li_2012,Yunes_2013,Sampson_2013,Agathos_2014,Abbott2016GRTests,Baker_2017,Isi_2017,isi_alan_2017,Callister_2017,Berti_2018,Takeda_2018,Abbott_2019BBH,Abbott_2021,Mehta_2023,sanger_2024,GWTC3_2025,Madekar_2025,kumar_2025,GWTC4_PTGR,Abac_2026}, and the planned and proposed next-generation GW detectors (e.g.\ Einstein Telescope~\cite{Punturo:2010zz}, Cosmic Explorer~\cite{Reitze:2019iox}, LISA~\cite{Shaddock_2009}) will improve the sensitivity of GW measurements by approximately an order of magnitude.

In GR, gravitational waves contain only two transverse tensor polarization modes: plus ($+$) and cross ($\times$) modes~\cite{Weinberg1972,MTW1973,Eardley_PRD,Will_1993,Maggiore2008,Schutz2009}. Modified metric theories of gravity, however, can allow up to six independent polarization modes. For example, scalar-tensor and massive-graviton models can give rise to additional breathing ($b$) and/or longitudinal polarization scalar modes, due to the presence of a scalar degree of freedom \cite{Will:2014kxa}. In contrast, vector-tensor gravity theories typically predict two longitudinal vector polarizations~\cite{Will:2014kxa}. More comprehensive theories like tensor-vector-scalar theories allow all six possible polarization states \cite{Bekenstein_2004,Bekenstein_2006}, whereas Einstein--Aether models support five distinct modes \cite{Jacobson_2004,eling2005,jacobson2008}. Similarly, bimetric theories are expected to exhibit the complete set of six polarizations~\cite{Will:2014kxa,Chatziioannou:2012rf,Lightman_73,ROSEN_1974}. Detection of non-tensorial polarization modes in GW signals would provide direct evidence that GR is incomplete.

Binary neutron star (BNS) mergers offer a distinct advantage for polarization tests: they produce electromagnetic (EM) counterparts that constrain the binary's orientation. The orientation is characterized by two angular parameters---the inclination angle $\iota$ and the polarization angle $\psi$---which are the primary quantities determining the polarization content of the GW signal in GR. GW170817~\cite{TheLIGOScientific:2017qsa} was the first multi-messenger event with comprehensive EM observations that enabled such measurements. The radio observations of the gamma-ray burst afterglow---characterized through Very Long Baseline Interferometry (VLBI)~\cite{Mooley:2018qfh,Ghirlanda:2018uyx}---provided constraints on both the inclination angle \cite{hotokezaka2018} and the polarization angle \cite{Lagos:2024boe}. The kilonova observation further provided sky localization and redshift measurements~\cite{2041-8205-848-2-L12}, which break parameter degeneracies inherent in GW-only analyses. Since EM counterparts are not typically expected from binary black hole mergers, BNS systems thus provide unique opportunities to improve polarization measurements through the combination of GW and EM data.

Most research to date has tested gravity through modifications to the amplitude and phase of GWs \cite{LIGOScientific:2021sio}, but not through polarization content. The few polarization tests performed use binary black hole / neutron star mergers, where parameter degeneracies result in poorly constrained polarizations \cite{Abbott_2017,Abbott:2017oio,Hagihara_2018,Abbott_2019,Abbott_2019BBH,Abbott_2021,GWTC3_2025,GWTC4_PTGR,Hagihara_2019,LIGOScientific:2021sio,Takeda2021_PolarizationTest,Svidzinsky2021_VectorGravity,Chatziioannou_2021,Takeda:2023wqn,Mehta_2023}. Some studies construct waveform models that allow multiple polarization modes (typically tensor and scalar) to coexist \cite{Takeda_2022}. In \cite{Takeda_2022}, accounting for scalar mode polarization induced energy loss up to second order yields corrections to both the amplitude and phase of the tensor modes, with these modifications parameterized by the scalar polarization parameter. The detector strain is written as a sum over polarizations weighted by antenna pattern functions, and Bayesian inference is used to estimate their amplitudes. This differs from pure polarization tests \cite{Takeda2021_PolarizationTest,Abbott:2017oio}, which assume one mode at a time and compare models; mixed models instead constrain non-tensor components while retaining the tensor contribution.

Parameterized tests of GR, such as the Flexible-Theory-Independent method \cite{Mehta_2023} and those in \cite{Abbott_2019BBH,Abbott_2021,GWTC3_2025,GWTC4_PTGR}, typically modify only the phase of the standard tensor ($+, \times$) modes, without introducing additional polarizations or amplitude changes. The deviations are applied across the inspiral and taper toward merger--ringdown; the leading $-1$PN parameter $\delta\phi_{-2}$ captures possible indirect dipole radiation. Other parameterized tests, such as \cite{Takeda_2022,Takeda:2023wqn}, do include extra polarizations with waveforms motivated by specific gravity theories. This is the approach that will be used in this paper.

Alternatively, in \cite{Hagihara_2018,Hagihara_2019,GWTC4_PTGR}, the null-stream method forms linear combinations of detector outputs that cancel the GR tensor modes, so any residual indicates non-tensor polarizations. This approach is model-independent and does not assume a specific waveform, but it does not use full alternative-polarization waveforms; residuals are treated phenomenologically, so constraints depend on detector geometry and do not capture detailed phase or amplitude effects to be able to map them to specific gravity theories.

\begin{figure}[h!]
	\centering
	\includegraphics[width = 0.49\textwidth]{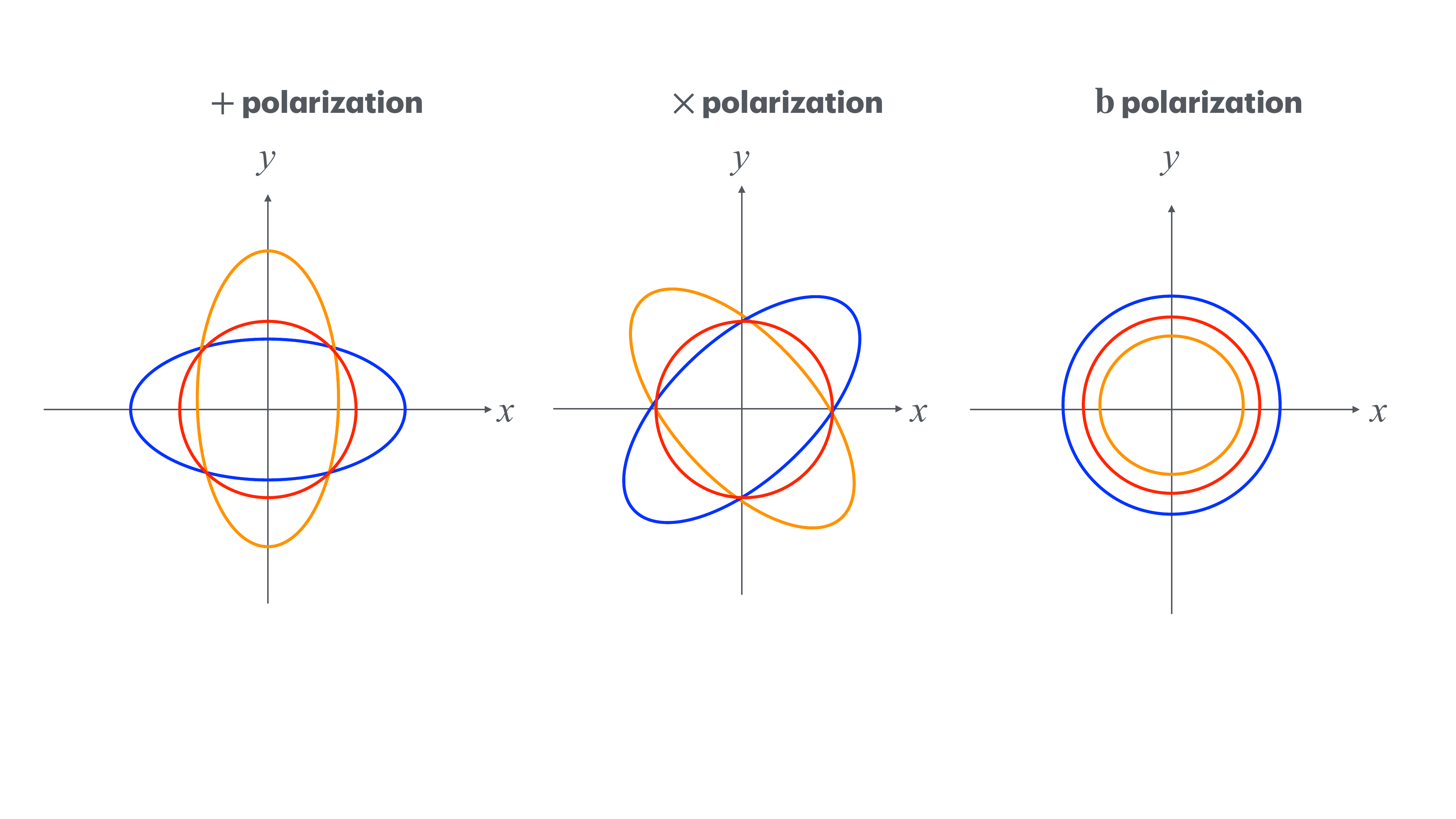}
	\caption{Illustration of how the polarizations $+$, $\times$ and b (scalar breathing) of a GW traveling towards $\hat{z}$ affect a ring of test particles. The red circle represents the unperturbed ring, while the orange and blue ellipses/circles represent the perturbed ring at different moments in time.}
	\label{fig1}
\end{figure}

In this paper, we use the angular information from EM counterparts---incorporating for the first time the polarization angle constraints from the gamma-ray burst afterglow---to improve polarization tests of gravity using GW170817. Note that at least five non-aligned differential-arm GW detectors are needed in order to resolve all five independent GW polarizations (as the two scalar modes are not distinguishable by these types of detectors), \cite{Eardley_PRL,Eardley_PRD}. Given this current limitation of GW detectors, and motivated by the most commonly studied scalar-tensor gravity theories, we focus on testing the existence of only a scalar breathing polarization, additionally including amplitude and phase modifications of the tensor polarizations for consistency. Fig.\ \ref{fig1} illustrates how space is distorted in the presence of the tensor and breathing polarizations. 

We use a theory-agnostic approach and use the parameterized post-Einsteinian (PPE) framework~\cite{Yunes_2009} for describing the deviations from GR during the inspiral, where we will add six extra non-GR waveform parameters. We model deviations from GR in both the quadrupole $\ell = m= 2$ and dipole $\ell = m= 1$ angular harmonics, with two possible frequency evolutions. We perform Bayesian inference on the GW170817 signal before the merger incorporating all the EM information through priors. For $\ell = m= 2$, we find mild evidence for a scalar polarization mode, along with deviations from the GR tensor polarizations. Instead, for $\ell = m= 1$  we find no deviation from GR. We analyze how much each EM constraint (sky location, luminosity distances, inclination angle, polarization angle) helps improve the breathing scalar mode. The polarization angle constraint places very tight bounds on non-GR parameters; for instance, in the case $\ell = |m| = 2$, the bound on the tensor amplitude modification parameter improves by roughly a factor of two.

The paper is structured as follows: Section \ref{Meth} describes the waveform model for the three polarization modes studied within the PPE framework, and describes the incorporation of EM information and methodology for parameter estimation. In Section \ref{Res} we present the resulting constraints on deviations from GR, and the impact of the EM information. In Section \ref{Conc} we provide a summary and conclusions of this work. Throughout this work, we use geometric units where $G = c = 1$.

%%%%%%%%%%%%%%%%%%%%%%%%%%%%%%%%%%
%METHODOLOGY
%%%%%%%%%%%%%%%%%%%%%%%%%%%%%%%%%%
\section{Methodology}\label{Meth}\label{WM}
This section is structured as follows. In Sec.\ \ref{subsec:waveform} we summarize the polarizations of gravitational waves and describe the PPE formalism for tensor and breathing modes. In Sec.\ \ref{subsec:EM} we summarize the EM counterparts to the event GW170817 and the parameter constraints it provides. In Sec.\ \ref{Prior} we summarize the methodology of the Bayesian parameter estimation performed here, and the priors used.

\subsection{Waveform Model}\label{subsec:waveform}
A generic metric theory of gravity can allow up to six independent polarization modes~\cite{Eardley_PRL,Will_1993}: plus ($+$), cross ($\times$), vector-$x$ ($x$), vector-$y$ ($y$), breathing ($b$), and longitudinal ($\ell$). For arbitrary polarization content, the response $h$ induced in a GW detector can be written as:
\bea
h = \sum_p F_p(\theta,\phi,\psi) h_p(t,\iota,\phi_c) ,
\eea
where the sum runs over the six polarizations \(p \in \{+, \times, x, y, b, \ell\}\) and the detector response functions \(F_p\) for an L-shaped detector are given by \cite{Sathyaprakash_2009} 
\begin{align}
F_{+} &= \tfrac{1}{2}(1+\cos^{2}\theta)\cos 2\psi \cos 2\phi 
        - \cos\theta \sin 2\psi \sin 2\phi, \\
F_{\times} &= \tfrac{1}{2}(1+\cos^{2}\theta)\sin 2\psi \cos 2\phi 
        + \cos\theta \cos 2\psi \sin 2\phi, \\
F_{x} &= -\sin\theta \left(\cos\theta \cos 2\phi \cos\psi 
        - \sin 2\phi \sin\psi \right), \\
F_{y} &= -\sin\theta \left(\cos\theta \cos 2\phi \sin\psi 
        + \sin 2\phi \cos\psi \right), \\
F_{b} &= -\tfrac{1}{2}\cos 2\phi \sin^{2}\theta, \\
F_{\ell} &= \tfrac{1}{2}\cos 2\phi \sin^{2}\theta,
\end{align}
where $\theta$ and $\phi$ denote the polar and azimuthal sky coordinates of the source, and $\psi$ is the polarization angle characterizing the orientation of the binary plane in the sky. 

With three operational detectors during GW170817 (two LIGO sites and Virgo)\footnote{The two LIGO detectors are nearly coaligned, providing approximately two independent measurements.}, we can constrain up to three independent polarization modes \cite{Aasi_2015,Acernese_2015,Aso_2013,Takeda_2018}. We focus on the two tensor modes $(+, \times)$ plus one scalar mode. We choose a scalar mode over vector modes because vector polarizations are less commonly predicted and less well-studied theoretically.

Among scalar modes, we include the breathing mode rather than the longitudinal mode. In networks of quadrupolar detectors with arm lengths smaller than the gravitational wavelength, breathing and longitudinal modes are completely degenerate \cite{Katerina_2012} (see the detector response functions above), making them indistinguishable in a model-independent analysis. The breathing mode appears generically in scalar-tensor theories, while the longitudinal mode arises only when the scalar degree of freedom is massive \cite{Eardley_PRL,Will_1993,Capozziello_2008}. We therefore test for the breathing mode, which represents the more general case. This degeneracy may not persist for next-generation detectors (Cosmic Explorer, Einstein Telescope, LISA) with different geometries or longer arm lengths \cite{Lee_2008,Chamberlin_2012, Akama:2026igu}.

For planar non-precessing binary systems, each GW polarizations can be decomposed into angular harmonics as~\cite{Akama:2026igu}:
\bea
h_p(t,\iota,\phi_c) = 
\mathrm{Re}\left(
\sum_{\ell,|m|} 
h_{p}^{(\ell, |m|)}(t)\,
Y_{p}^{(\ell, |m|)}(\iota,\phi_c),
\right)
\eea
as functions of the inclination angle $\iota$ (characterizing the inclination of the binary plane with respect to the line of sight) and the coalescence phase $\phi_c$.
Since we will be considering a  nearly equal-mass binary, the dominant angular harmonic in the tensor polarizations is the quadrupole  $(\ell, |m|) = (2,2)$. Furthermore, the scalar polarization is expected to also emit radiation in the dipole angular harmonic $(\ell, |m|) = (1,1)$~\cite{Will_1993,Damour_1996}. For these cases of interest, the angular harmonics $Y_p^{(\ell,|m|)}$ are explicitly given by \cite{Akama:2026igu}
\bea
Y_{+}^{(2,2)}(\iota,\phi_c) =
\frac{1}{8}\sqrt{\frac{5}{\pi}}\,
\left(1+\cos^2\iota\right)
e^{2 i \phi_c},
\eea

\bea
Y_{\times}^{(2,2)}(\iota,\phi_c) = \frac{i}{4}\sqrt{\frac{5}{\pi}} \cos\iota e^{2 i \phi_c},
\eea
\bea
Y_{b}^{(2,2)}(\iota,\phi_c) =
\frac{1}{4}\sqrt{\frac{15}{2\pi}}\,
\sin^2\iota
e^{2 i \phi_c},
\eea

\bea
Y_{b}^{(1,1)}(\iota,\phi_c) =
-\frac{1}{2}\sqrt{\frac{3}{2\pi}}\,
\sin\iota
e^{i \phi_c}.
\eea

Due to the fact that GWs are generally weak, the signals are typically detected via a match filtering procedure, where assumptions about the signal morphology are made and compared to the data \cite{Abbott:2016blz}. We will use the so-called Parameterized Post-Einsteinian (PPE) formalism \cite{Yunes_2009} to describe the modifications to GR in a theory-agnostic way. This framework incorporates power-law frequency-dependent terms in the amplitude and phase of the GW waveform motivated by a Post-Newtonian (PN) expansion during the inspiral. Examples of gravity theories mapped to the PPE framework and applications can be found in e.g.\ \cite{Tahura:2018zuq, Perkins:2020tra, Carson:2020rea, Xie:2024ubm}.

In this paper, we will introduce modifications to the amplitude and phase of the two tensor modes, as done in the previous literature.
Typically, the PPE formalism is used to describe modifications of the dominant $(\ell=2,|m|=2)$ harmonic, neglecting higher angular harmonics~\cite{Yunes_2009}, as we will assume here. In this case, the tensor polarizations in frequency domain are modified via power-law corrections as:
\begin{eqnarray}
    \tilde{h}_{+/\times}^{(2,2)}(f)=  \tilde{h}_{\rm GR,+/\times}^{(2,2)}(f) (1 + \alpha u_2^{a_T})e^{i\beta u_2^{b}}, \label{Eq:ppE_original}
\end{eqnarray}
where $\tilde{h}(f)^{(\ell,|m|)}$ is the tensor mode in frequency-domain obtained by a Fourier transform as :
\[
\tilde{h}_p^{(\ell,|m|)}(f) = \int_{-\infty}^{\infty} h_p(t)^{(\ell,|m|)}(t)\, e^{2\pi i f t} \, dt.
\]
We have also introduced the dimensionless factor $u_m$ which is generally defined for any angular harmonic as:
\begin{eqnarray}
    u_{m}=\left(\frac{2\pi \mathcal{M}_zf}{|m|}\right)^{1/3}. \label{eq:ul-PPE-extra}
\end{eqnarray}
Here, \(\mathcal{M}_z = (1 + z)\mathcal{M} = \eta^{3/5}M(1+z)\) is the redshifted chirp mass, with $M = m_1 + m_2$ being the total mass of the binary and $\eta = m_1 m_2/M^2$ the symmetric mass ratio. 
Furthermore, we have introduced the parameters ($\alpha, a_T$) that modify the tensor amplitude, and  ($\beta, b$) that modify the phase. These four new parameters will be assumed to be constant and real, and describe the leading-order PN corrections to GR. When $\alpha=\beta=0$ we recover the tensor polarization in GR, and any non-zero values will indicate a modification to GR.

Next, based on the work in \cite{Chatziioannou:2012rf, Akama:2026igu}, we propose the following model for the breathing polarization for both quadrupole and dipole angular harmonics:
\begin{align}
    \tilde{h}_{b}^{(\ell,|m|)}(f)  = A_{PN} \alpha_{B} u_m^{a_{S}}e^{-i\Psi^{(m)}_{GR}} e^{i m\beta u_m^{b}/2}, \label{Eq:h_PN}
\end{align}
where the amplitude \(A_{PN}\) is given by the inspiral GR waveform as \cite{Khan_2016}
\begin{eqnarray} 
A_{PN}(f) = A^{(m)} \sum_{i=0}^{6} A_i \left(\frac{2\pi f}{|m|}\right)^{i/3},
\label{Eq:A_pn_scalar}
\end{eqnarray}
with $A_i$ being the PN expansion coefficient (see the Appendix section B of ~\cite{Khan_2016}), and $A^{(m)}$ being the leading PN amplitude factor in GR
\begin{align}
    A^{(m)}=\frac{2\pi}{(3|m|)^{1/2}}\frac{\mathcal{M}_z^2}{d_L} \eta^{(2-|m|)/5} u_m^{(2|m|-11)/2}, \label{Eq:h_PN_aell}
\end{align}
where $d_L$ is the source luminosity distance. In Eq.\ (\ref{Eq:h_PN}) the new PPE parameter $\alpha_B$ is introduced to describe the magnitude of the scalar amplitude, and $a_S$ to describe the deviation on the frequency evolution compared to GR. Both $(\alpha_B,a_S)$ are assumed to be real constants. The phase in Eq.~(\ref{Eq:h_PN}) also comes from the inspiral GR phase, which is explicitly given by \cite{Buonanno_2009} 
\begin{eqnarray}
\Psi^{(m)}_{GR} = - 2 \pi ft_c + \frac{\pi}{4} - \frac{3 |m|}{256 u_{m}^5} \sum_{n = 0}^7 u_{m}^{n} \left( c_n^{PN} + l_n^{PN} \ln u_{m} \right)\label{Psi_GR}
\end{eqnarray}
Here $c_n,l_n$ are PN expansion coefficients that are
functions of the intrinsic binary parameters (See Appendix B of ~\cite{Khan_2016}),
and the parameters $(\beta,b)$ describe the phase deviations compared to GR. Notice that the same PPE parameters are used in the phase of the tensor and scalar polarizations, since this is found to be the case in various modified gravity theories \cite{Chatziioannou:2012rf, Akama:2026igu}.

The PPE formalism was originally derived to describe leading-order deviations from GR during the inspiral phase ~\cite{Yunes_2009}. Many recent works \cite{Abbott_2021,Mehta_2023,Schumacher_2023,GWTC3_2025,Piarulli_2025, Akama:2026igu} have extended the PPE modifications through the merger and ringdown phases by applying a taper function that smoothly removes the non-GR corrections as the binary approaches merger. This approach implicitly assumes that deviations from GR become negligible during merger and ringdown---a strong assumption without clear theoretical justification, yet commonly used to minimize parameter degeneracies. We take a more agnostic approach and apply the PPE framework only to the inspiral portion of the waveform, terminating it at a cut-off frequency $f^{\ell m}_{\rm c}$ before the merger begins. This ensures our analysis tests deviations during the phase for which the PPE framework was designed, without making assumptions about the behavior of modified gravity during the highly nonlinear merger and ringdown.

%%%%%%%%%%%%%%%%%%%%%%%%%%%%%%%%%%
\subsection{EM counterparts}\label{subsec:EM}
The gravitational-wave event GW170817, detected by the LIGO and Virgo detectors on 17 August 2017, was the first confirmed binary neutron star merger observed through both gravitational waves and electromagnetic radiation \citep{Abbott_2017,Abbott2017_L13}. A kilonova is sufficient to enable a three-dimensional localization of the source. 
Its position on the sky is determined by the coordinates $(\mathrm{ra}, \mathrm{dec})$, 
while its distance $d_L$ or redshift $z$ can be estimated—particularly when the host galaxy is identified. 
In addition, a gamma-ray burst and its afterglow allow the orientation of the binary plane to be inferred through the angles $(\iota, \psi)$. In the case of GW170817, approximately $1.74\,\mathrm{s}$ after the GW merger, a short gamma-ray burst GRB~170817A was detected by \textit{Fermi}-GBM and \textit{INTEGRAL}, providing the first direct evidence linking short gamma-ray bursts to neutron-star mergers \citep{Goldstein2017,Savchenko2017,LIGOScientific:2017zic}. Optical follow-up observations subsequently identified the kilonova AT2017gfo, whose rapidly evolving blue-to-red emission is consistent with radioactive heating from r-process nucleosynthesis in neutron-rich ejecta \citep{SoaresSantos2017,Kasen2017}. Multi-wavelength observations further revealed delayed X-ray and radio afterglow emission consistent with a structured relativistic jet viewed off-axis \citep{Margutti2018,Mooley:2018qfh}. When the viewing angle exceeds the jet half-opening angle, the system is observed off-axis. 

Thanks to the kilonova, GW170817 was localized at $(\mathrm{ra},\,\mathrm{dec}) = (13{:}09{:}48.085 \pm 0.018 s,\,-23{:}22{:}53.343 \pm 0.218'')$ to the galaxy NGC~4993 at redshift $z \simeq 0.0098$, corresponding to a luminosity distance $D_L \approx 42.9\pm3.2\,\mathrm{Mpc}$~\cite{LIGOScientific:2017zic}.

High-angular-resolution VLBI radio observations allows to follow-up the GRB afterglow for months to years after the merger. These radio observations allow us to infer both the viewing angle\footnote{The viewing angle corresponds to the inclination angle $\iota$ for a face-on binary configuration, or to $\pi - \iota$ for a face-off configuration.} and the polarization angle. The inclination measurement relies on detailed modeling of the jet structure and its light curve~\cite{Finstad2018,2041-8205-848-2-L12}, whereas the polarization angle can be determined more directly by tracking the jet’s proper motion and trajectory across the sky~\cite{Mooley:2018qfh}.

From EM observations, the polarization angle of GW170817 is constrained to  $\psi = 3.14 \pm  0.09\, \mathrm{rad}$ at $68\%$ credibility ~\cite{lagos_2024_btgr}, and the binary inclination to $\iota$= 2.85 \(\pm\) 0.03 rad \citep{hotokezaka2018}.

%%%%%%%%%%%%%%%%%%%%%%%%%%%%%%%%%%
\subsection{Parameter Estimation}\label{Prior}

For waveform generation and parameter estimation, we use \texttt{Jim} \cite{Edwards:2023sak,Wong:2023lgb,Wouters:2024oxj}, which implements the IMRPhenomD waveform, a quasi-circular binary black hole non-precessing waveform model \cite{Khan_2016}. The term $\tilde{h}^{(2,2)}$\(_{\textrm {GR},+/\times}\) in Eq.\ (\ref{Eq:ppE_original}) will then be given by the IMRPhenomD waveform. We apply it to the binary neutron star GW170817, neglecting tidal effects as tidal contributions become significant only at late inspiral stages and are not expected to be degenerate with the PPE modifications considered here. For comparison we also performed an analysis with \texttt{IMRPhenomD\_NRTidalv2} waveform~\cite{Dietrich_2017,Dietrich_2019} and presented the result in Appendix \ref{Appendix_D}. 

In addition, the cut-off frequency between the inspiral and intermediate regions for \((\ell,|m|\))=(2,2) will be defined as \cite{Abbott_2021,Mukherjee24}\
\[
    M_z f^{22}_{\rm c} = 0.018
\]
where \(M_z=M(1+z)\) is the total redshifted mass of the binary. The cut-off frequency \(f^{\ell m}_{\rm c}\) will be obtained by using $M_z =  2.78 M_\odot$  which is the best-fit value from the parameter estimation analysis of the entire signal of GW170817  (i.e.\ inspiral, merger, and ringdown) using a GR waveform template. Since different angular harmonics contain GWs at different frequencies, the cut-off frequency for \((\ell,|m|\))=(1,1) mode will be given by \(f^{11}_{\rm c} = f^{22}_{\rm c}\)/2. The cut-off frequency for GW170817 in this analysis is \(f^{22}_{\rm c}\approx 1300 \)Hz.  Notice that the choice of cut-off frequency is not unique. For instance, in \cite{Takeda_2018} the frequency is taken to be $f_{\mathrm{ISCO}}$, equal to twice the orbital frequency of the innermost stable circular orbit for a point mass in Schwarzschild spacetime. In this work, we varied the cut-off frequency to test the robustness of our results and found similar results (see Fig.\ \ref{fig8}) in Appendix \ref{Appendix_C}).

We test for deviations from GR by introducing a scalar breathing polarization mode into the gravitational waveform alongside modifications to the tensor modes. We employ the PPE framework, which supplements the standard 11 GR parameters of a non-precessing compact binary (chirp mass $M_c$, mass ratio $q$, inclination $\iota$, sky position $(\mathrm{ra}, \mathrm{dec})$, polarization angle $\psi$, coalescence time $t_c$ and phase $\phi_c$, luminosity distance $d_L$, and aligned spin components ($s_{1z}$, $s_{2z}$) with six non-GR parameters. Three exponents $(a_T, b, a_S)$ characterize the frequency dependence of the deviations, and three amplitudes $(\alpha, \beta, \alpha_B)$ quantify their strength. In our Bayesian analysis, we sample over the 11 GR and the 3 PPE parameters, while fixing the exponents to two sets of values motivated by specific modified gravity theories (see Table (\ref{tab1})). To date, calculations of extra polarization waveforms in modified gravity have been performed only for a limited set of theories, including Horndeski gravity~\cite{Higashino_2023}, Einstein-\ae ther theory~\cite{Zhang:2019iim}, Rosen’s theory~\cite{ROSEN_1974455}, and Lightman--Lee theory~\cite{Lightman_73}. The PPE framework maps these theories~\cite{Akama:2026igu} into a common parameterization in which the frequency scaling of tensor and additional scalar polarizations (breathing mode) is characterized by $(a_T, b, a_S)$ with only a few discrete values taken, depending on the dominant coupling parameters. In Horndeski gravity, two regimes arise, $(a_T,b,a_S)=(-2,-7,0)$ and $(0,-5,0)$, corresponding to the leading and sub-leading PN deviations from GR, respectively. In Einstein-\ae ther theory, similar values of $a_T$ and $b$ occur, but $a_S$ can also be nonzero (-2). In particular, $b=-7$ modifies the $-1$PN contribution to the GR phase, which will be the main focus of this paper and hence we do not consider $b=-5$.

\begin{table}[h!]
\centering

\begin{tabular}{|c@{\hskip 8pt}|ccc|ccc|}
\hline
%\toprule
\textbf{Set} & \multicolumn{3}{c|}{\textbf{Fixed}} & \multicolumn{3}{c|}{\textbf{Free}} \\
            & $a_T$ & $b$ & $a_S$ & $\alpha$ & $\beta$ & $\alpha_B$ \\
            \hline
\midrule
$1$ & $-2$ & $-7$ & $-2$ & [$-10^{-2}$, $10^{-2}$] & [$-10^{-4}$, $10^{-4}$] & [$-5$, $5$] \\
$2$ & $-2$ & $-7$ & $0$ & [$-10^{-2}$, $10^{-2}$] & [$-10^{-4}$, $10^{-4}$] & [$-35$, $35$] \\
\hline
\end{tabular}
\caption{Choices of the fixed PPE parameters \( a_T \), \( b \) and \( a_S \), and prior distributions for $(\alpha,\beta,\alpha_B)$. We use uniform priors between a minimum (`min') and a maximum (`max'), denoted as [min,max]. Here $(\alpha,\alpha_B)$ are dimensionless but $\beta$ is in the unit of radian. The quoted ranges of these parameters are for the $\ell=|m|=2$ case and we use the same priors for $\ell=|m|=1$.}\label{tab1}
\end{table}

In the PPE formalism it is assumed that leading-order PN deviations from GR are perturbative. While, the priors for GR parameters will follow Ref.~\cite{Wong:2023lgb}, the non-GR PPE parameters are constrained by consistency bounds from perturbation theory: $\alpha u_2^{a_T} \leq 1$, $|\beta u_2^{b}| \leq \pi$, and $\alpha_B u_2^{a_S}\sin^2(\iota) \leq 1$ for $\ell=|m|=2$ or $\alpha_B u_1^{a_S}\sin(\iota) \leq 1$ for $\ell=|m|=1$, where $u_m$ is defined in Eq.~(\ref{eq:ul-PPE-extra}). In order to be conservative, these bounds are evaluated at the minimum frequency $f=23$ Hz and the best-fit inclination angle coming from the EM data $\iota=2.85$ rad, with the resulting ranges in the third column of Table \ref{tab1}.
Here we assume that the amplitudes of the scalar polarization is small compared to the dominant $(\ell = |m| = 2)$ tensor mode in GR. We denote it as ``GW only'' case where all the 11 GR  and 3 non-GR PPE parameters are uniformly distributed in their relevant range. 

\begin{figure*}[t]
    \centering
    \includegraphics[width=0.9\textwidth]{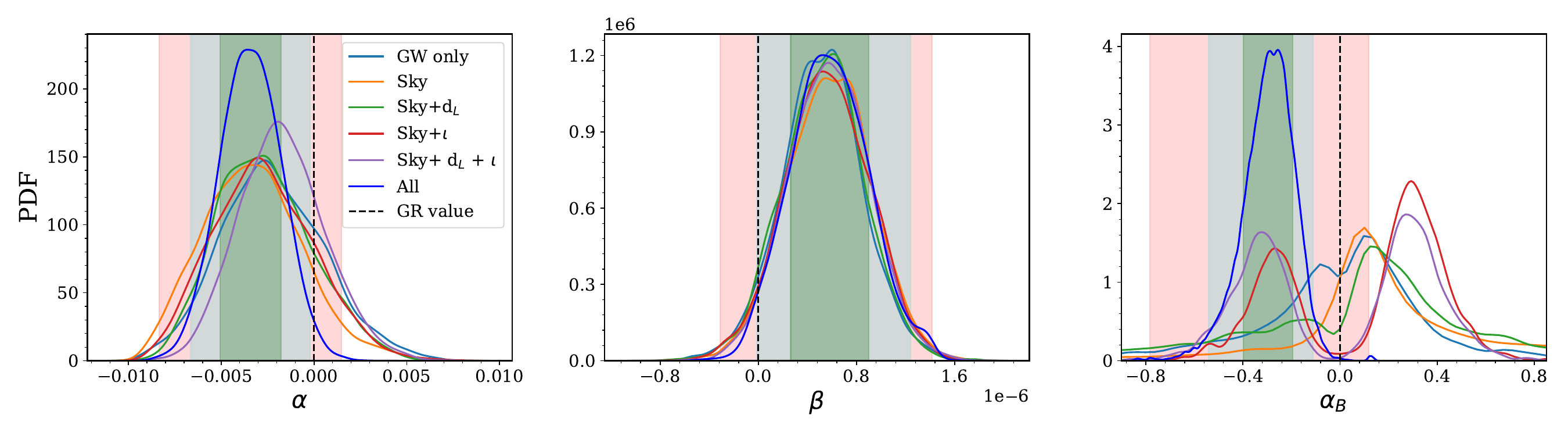}
    \caption{Posterior distribution function (PDF) of the three PPE parameters for various EM priors, as shown in colors, for the quadrupole $\ell = |m|= 2$,  fixing $a_T = a_S = -2$ and $b = -7$. The vertical bands show the 68\% (green), 95\% (cyan), and 99.7\% (red) credible intervals (CI) for the “All” case. The black dashed line represents the parameter values predicted by General Relativity (GR). }
    \label{fig2}
\end{figure*}

In order to assess the impact of electromagnetic constraints on the PPE parameters, we perform the analysis with progressively more restrictive priors, starting from a baseline case with uniform priors on all GR parameters (denoted (i) ``GW only''). The cases included will be: (ii) ``Sky'': Gaussian priors (\(\mu\pm\sigma\)) with a mean \(\mu\) and standard deviation \(\sigma\) on the sky localization parameters, i.e., ra = 3.44 $\pm$ 0.01 rad and dec = -0.40 $\pm$ 0.01 rad. In principle, the true uncertainty is very small (see Sec.~\ref{subsec:EM}, the uncertainties, when converted to radians, are of order $10^{-6}\,\mathrm{rad}$.), but to avoid numerical issues in the computation, we use this value. This choice does not influence the results, as demonstrated later in Fig.~\ref{fig1}. (iii) ``Sky + $d_L$'', \( d_L \)= 42.9 \(\pm\) 3.2 Mpc. (iv) ``Sky + $\iota$'', \( \iota \)= 2.85 \(\pm\) 0.03 rad. (v) ``Sky + \( d_L \) + \( \iota \)'' combining the priors of the previous cases. (vi) ``All'', where all previous priors are used together with the polarization angle \( \psi \)= 3.14 \(\pm\) 0.09 rad.

%%%%%%%%%%%%%%%%%%%%%%%%%
%RESULTS
%%%%%%%%%%%%%%%%%%%%%%%%%
\section{Results}\label{Res}
In this section we present the main results from Bayesian parameter estimation, focusing on the three PPE parameters $(\alpha,\beta,\alpha_B)$. In Sec.\ \ref{sec:quadrupole} we show the results when the breathing mode has a quadrupolar signal, and in Sec.\ \ref{sec:dipole} when it has a dipolar signal. Recall that in all cases the tensor mode only has quadrupolar signal, as in GR.

\subsection{Quadrupole angular harmonic: $\ell = |m|=2$}\label{sec:quadrupole}

\subsubsection{Frequency evolution: $a_T=-2$, $b=-7$, $a_S=-2$}
We first analyzed the most studied parameter set in the literature, fixing \( a_T = -2 \), \( b = -7 \), and \( a_S = -2 \).
Parameter estimation was initially performed using only the GW data (case ``GW only''), and then refined by incorporating priors from the EM observations. 

Fig.\ \ref{fig2} shows the posterior distributions of the three PPE parameters $(\alpha,\beta,\alpha_B)$ for various prior information as discussed in Sec~\ref{Prior}, with the dark blue color showing the best constraints for the case ``All''. In the leftmost panel of Fig.\ \ref{fig2}, the posterior distributions of the parameter \( \alpha \) are shown. In the case ``All" there is a strong bound on $\alpha = -3.43^{+3.21}_{-3.21} \times 10^{-3}$ at 95\% credible interval (CI). The vertical color bands show the $68\%,\, 95\%,\, 99\%$ CI for the  ``All'' case. We find that the posteriors for the cases ``GW only'', ``Sky'', ``Sky+$d_L$'' and ``Sky+$\iota$''
are quite similar,  $\approx -2.84^{+5.58}_{-4.69} \times 10^{-3}$ at 95$\%$ CI, suggesting that these pieces of EM information do not help improve constraints on $\alpha$. However, the uncertainties are reduced by nearly  $12\%$ in the case ``Sky+$d_L$+$\iota$''\footnote{Notice that the central value shifts in the ``Sky+$d_L$+$\iota$'' case, which is due to the known fact that the EM constraints for $d_L$ and $\iota$ do not coincide with the best-fit values from the GW data.}, and further improved by $29\%$ in the ``All'' case, which shows that the inclusion of $\psi$ helps improve the constraints on $\alpha$. In this latter case, we find GR (i.e.\ $\alpha=0$) to lie at the $\sim$ 96.4$\%$ CI of the posterior.

For the ``All'' case, the full corner plot for the $11+3$ parameters is shown in Appendix~\ref{Appendix_G} (see Fig.\ \ref{fig12}). We see that \(\alpha\) has a moderate positive correlation  with the luminosity distance, \(r[\alpha-d_L]\approx 0.7\), where  \(r[A-B]\) denotes the Pearson correlation coefficient between A and B. This happens because the response scales as  \(h \propto (1+\alpha)/d_L\) and hence an increase in \(\alpha\) can be compensated by increase in \(d_L\). In all other cases, except ``All'', this correlation goes away as $d_L$ and/or \(\alpha\) are allowed to correlate with other free parameters such as \(\iota,\, \psi\) and \(\alpha_B\). 

In the middle panel of Fig.\ \ref{fig2},  the posterior distributions of the parameter  \(\beta\) for the same cases are presented. We overall find that the posterior is unchanged when adding EM information since EM observations constrain waveform parameters that mostly affect the amplitude of GWs (three angles in the antenna pattern function, luminosity distance, and inclination in the angular harmonic functions). In the ``All'' case, we obtain $\beta=(5.77^{+6.65}_{-5.90} \times 10^{-7}) \times 10^{-7}$ rad at $94.1\%$ CI, which is consistent with the GR value $\beta=0$.

As seen in the full corner plot in Appendix \ref{Appendix_G} (see Fig.\ \ref{fig12}), a very strong anti-correlation between \(\beta\) and \(\mathcal{M}\) is found for all the scenarios considered, with \(r[{\beta-{\mathcal M}}]\approx -1.0\)  in the ``All'' case. This correlation makes sense since the total phase of tensor modes can be written as the Fourier-domain phase (leading order) in the PPE framework can be written as
\begin{equation}
\Psi(f) = - 2 \pi f t_c + \frac{\pi}{4} - \frac{3}{128} u_2^{-5} + 2\phi_c + \beta u_2^{-7},
\end{equation}
Although the PPE correction $\beta u_2^{-7}$ is subdominant compared to the leading GR term $-u^{-5}$, we observe a strong negative correlation between the chirp mass ${\mathcal M}$ and the PPE parameter $\beta$. This is because parameter estimation is sensitive to variations of the phase rather than the absolute magnitude of individual contributions. Considering small parameter variations, the phase change can be expressed as
\begin{equation}
\delta \Psi \approx 
\frac{\partial \Psi}{\partial {\mathcal M}}\,\delta {\mathcal M} +
\frac{\partial \Psi}{\partial \beta}\,\delta \beta.
\end{equation}
Along directions where the waveform remains approximately unchanged ($\delta \Psi \approx 0$), this leads to
\begin{equation}
\frac{\partial \Psi}{\partial {\mathcal M}}\,\delta {\mathcal M} +
\frac{\partial \Psi}{\partial \beta}\,\delta \beta \approx 0,
\end{equation}
with \(\frac{\partial \Psi}{\partial {\mathcal M}}=\frac{5}{128}(\pi(1+z)f)^{-5/3}{\mathcal{M}}^{-8/3} -\frac{7}{3}\beta(\pi(1+z)f)^{-7/3}{\mathcal{M}}^{-10/3}>0\) as the first term dominate and \(\frac{\partial \Psi}{\partial \beta}>0\)
which implies
\begin{equation}
\delta \beta \propto -\,\delta {\mathcal M}.
\end{equation}
This degeneracy arises because variations in ${\mathcal M}$ alter the phase evolution in a manner that partially mimics the frequency dependence of the PPE correction. As a result, even a small contribution such as $\beta u^{-7}$ can exhibit a strong correlation with ${\mathcal M}$ over the finite detector bandwidth.

In the rightmost panel of  Fig.\ \ref{fig2}, we show the posterior distributions of the parameter \(\alpha_B\), which determines the scalar breathing amplitude. A non-zero value for \(\alpha_B\) would suggest the existence of a scalar breathing mode in the signal. We generally find that the distribution tends to be extended and double peaked.
\begin{figure*}[t]
\includegraphics[width=0.45\textwidth]{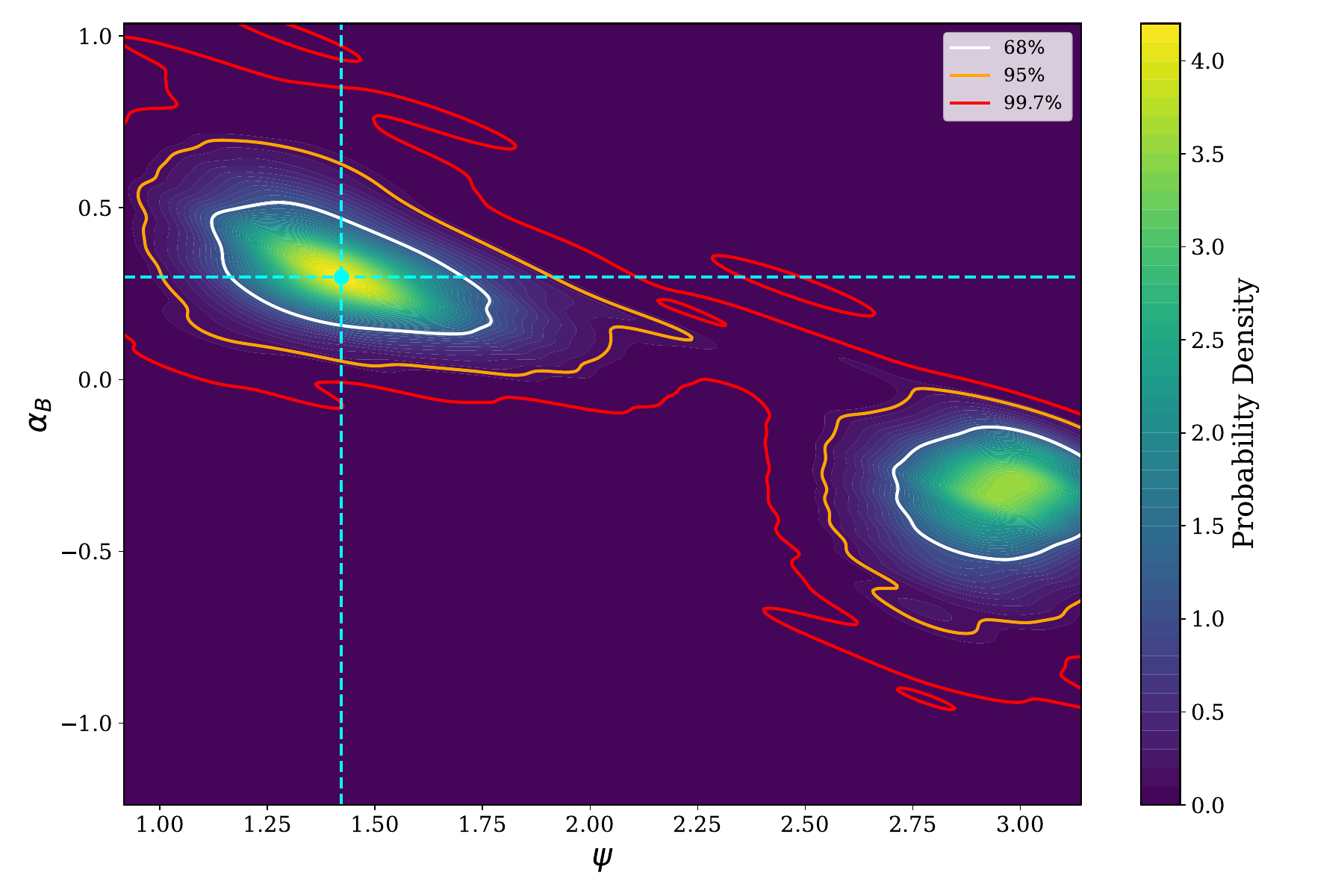} 
\includegraphics[width=0.45\textwidth]{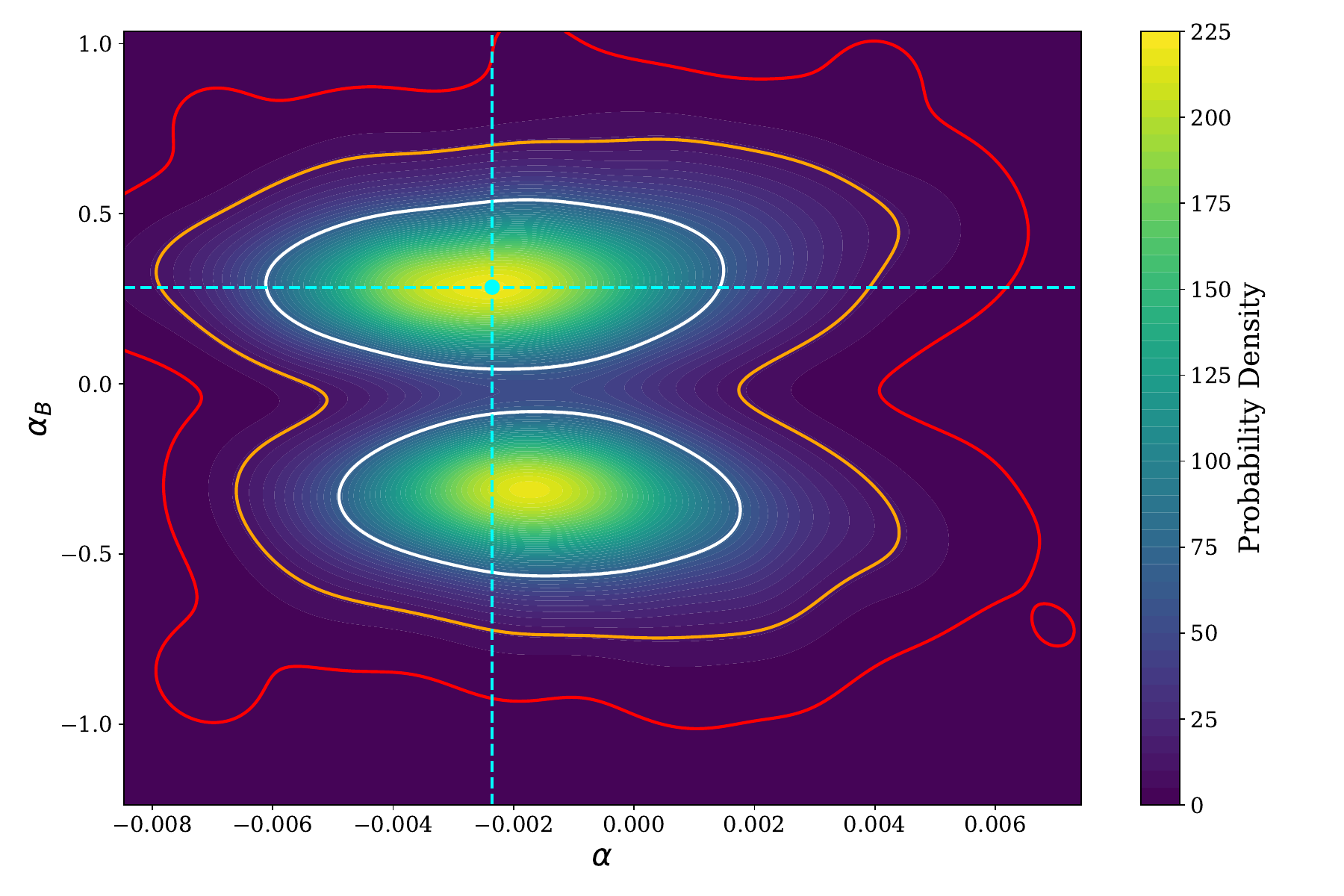} \\
\caption{Joint posterior distributions for $\alpha_B-\psi$ (left plot)  and $\alpha-\alpha_B$ (right plot), in the case ``Sky + d\(_L\) + \(\iota\)". These plots are for $\ell = |m|$ = 2 keeping $a_T = a_S = -2$ and $b = -7$. The green dots, where the dashed lines meet, show the best-fit values.}
    \label{fig3}
\end{figure*}
 The double peak arises because the IMRPhenomD waveform has a discrete symmetry $\psi\rightarrow\psi + \pi/2$~\cite{Roulet_2022} for (2,2)-harmonic
dominated waveforms for tensor polarizations. Under this discrete shift in $\psi$, both $h_+$ and $h_\times$ flip sign and so the angular pattern functions $F_+,F_\times$. A corresponding flip in the sign of $\alpha_B$ in the breathing mode can thus compensate for the change in $\psi$ (see more details in Appendix \ref{Appendix_A}). When  \(\psi\) is constrained by the EM observations in the ``All" case, the double peaks in \(\alpha_B\) disappear, as seen in the Fig.\ \ref{fig2}. The resulting best constraint is then given by $ \alpha_B\approx-0.29^{+0.18}_{-0.26}$ at $95\%$ CI, which has the GR value $\alpha_B=0$ lies approximately at the \(\sim 99.75\%\)CI of the posterior distribution. These results show that the tighter constraint on \( \alpha_B \) in the ``All'' case arises from restricting \(\psi\) and breaking its $\pi/2$ degeneracy. The transition from the double-peaked posterior to a single-peaked posterior is therefore driven by the inclusion of the polarization-angle prior. This prior is not an arbitrary statistical assumption, but is derived independently from electromagnetic observations of the GRB afterglow and it is thus data driven. Consequently, the single-peaked posterior reflects the combination of the GW data with independent EM information that breaks the intrinsic $\alpha_B$--$\psi$ degeneracy. This shows the important role that constraints on polarization angle have in constraining extra polarizations of gravitational waves.  The inclusion of polarization angle constraint improves the bound on $\alpha_B$ by $\approx 61\%$ compared to the double peaks in the ``Sky+$d_L$+$\iota$'' case.

Figure~\ref{fig3} shows contour plots of $\alpha_B-\psi$ (left) and $\alpha-\alpha_B$ (right) for the ``Sky + $d_L$ + $\iota$'' case. Splitting the data by $\alpha_B<0$ and $\alpha_B>0$, we find a consistent correlation $r[\alpha-\psi]\approx -0.7$ in both subsets. In contrast, $r[\alpha_B-\psi]\approx 0.27$ for the left peak and $\approx -0.44$ for the right peak. When $\alpha_B$ is restricted to a single (left) peak by fixing $\psi$ in the ``All" case, the correlation reduces to $r[\alpha_B-\psi]\approx 0.2$. 

Overall, in the inspiral test we have performed here, we are finding about a $2-3\sigma$ preference for modified gravity. As a comparison, in \cite{Takeda_2022} the authors extend gravitational-wave models to include a possible scalar polarization and constrain the scalar-to-tensor amplitude ratio using Bayesian inference on some GW events. For GW170817, they place a stringent upper bound on the scalar-to-tensor ratio ($\lesssim 0.068$) and on the additional scalar polarization parameter  \(\sim
0.04^{+0.60}_{-0.66}\)
at 68\% confidence level, finding it consistent with GR. Overall, no statistically significant deviation from General Relativity is observed, and the results show no meaningful preference for beyond-GR polarization content. Compared to their analysis, we adopt a narrower prior on $\iota$ and include a prior on the polarization angle. 
The amplitude of the scalar mode for GW170817 differs from ours (see Eq.~(\ref{Eq:A_pn_scalar}) and Eq.~3 of~\cite{Takeda_2022}), 
and we terminate the waveform above the cut-off frequency. Also in~\cite{Takeda_2022} , the inclusion of scalar-mode–induced energy loss up to second order leads to corrections in both the amplitude and phase of tensor modes, parameterized by the scalar polarization parameter.
These differences may lead to variations in the results.

Whether this mild preference for modified gravity reflects a genuine physical signal, or arises from statistical fluctuations in the detector noise~\cite{Svidzinsky2021_VectorGravity}, or model assumptions on the EM posteriors obtained for $\iota$ and $\psi$, remains an open question. It is worth noting that similar tensions have appeared in other analyses. Parameterized tests and ringdown studies reported in \cite{gwtc4_tgr1,GWTC4_PTGR,gwtc4_tgr3} have found several individual events---including GW230628\_231200, GW231028\_153006, GW231110\_040320, and GW231123---whose inferred parameters deviate from GR predictions beyond the $90\%$ credible interval.  Likewise, both the TIGER and PCA analyses of GW190814 \cite{Abbott_2020} yield deviations from GR outside the $90\%$ CI. However, if the underlying gravitational physics truly departed from GR, such deviations would be expected to appear consistently across all binary black hole (BBH) events. When the full population of BBH events in GWTC-4 is analyzed jointly, no statistically significant deviation from GR emerges, indicating that the outliers observed in individual events are most likely attributable to noise fluctuations.

Moreover, the argument that deviations from GR should appear uniformly across all compact binary mergers implicitly assumes that modified gravity effects are source-independent. However, this need not be the case. Black holes in several modified gravity theories satisfy no-hair theorems \cite{Israel_1967,Carter_1971} that suppress or entirely eliminate scalar charges, whereas neutron stars can develop significant scalar configurations depending on their equation of state and compactness (e.g.\ through spontaneous scalarization \cite{Silva2018_SpontScalarization,Doneva2024_SpontScalarization}). Consequently, binary neutron star mergers may exhibit deviations from GR that are absent in binary black hole systems. If this is the case, stacking BBH events to test for departures from GR would dilute or mask any signal that is specific to BNS systems. The mild preference for a scalar breathing mode found in our analysis of GW170817 could therefore reflect genuine source-dependent beyond-GR physics that would only be detectable through dedicated analyses of BNS events. Whenever the number of observed BNS mergers with electromagnetic counterparts grows, it will become possible to test this hypothesis by performing further analyses restricted to neutron star systems.

Finally, we investigate the impact of testing one given PPE parameter at a time (`pure' case) versus the joint PE for the three parameters (`mixed' case) for the “All” case (see Appendix~\ref{Appendix_B}). When tensor and scalar amplitude modifications are included simultaneously, the distributions shift away from the GR values. In contrast, modifying only the tensor amplitude and phase (i.e.\ a model with a vanishing scalar polarization) shows no significant preference for modified gravity, indicating that any mild preference arises only when all three parameters are modified together. This is important to take into consideration since many tests of gravity with GWs only consider modifications in the tensor polarizations~\cite{Mehta_2023}.

A polarization test was performed in~\cite{Takeda2021_PolarizationTest} on GW170817 and GW170814, considering pure tensor, vector, and scalar polarizations. Bayesian model selection was employed to compute Bayes factors between the different polarization hypotheses. For GW170817, incorporating sky-location and jet-orientation priors further reinforced support for tensor modes, strongly disfavoring scalar and vector polarizations. Overall, their results provide robust evidence for GR tensor polarizations. In that analysis, all non-tensorial polarizations were modeled using the same waveform as the GR tensor mode, differing only in their inclination dependence. Similar conclusion is drawn in \cite{Abbott:2017oio}. In contrast, our analysis employs a distinct waveform for the scalar breathing mode and introduces phase modifications for all three polarization modes, which may explain the differences in results compared to theirs. We cannot directly compare our results with pure polarization tests, as we do not encounter analogous scenarios where all of $\alpha$, $\beta$, and $\alpha_B$ vanish, with at least one of these parameters always being present.

In Ref.~\cite{Mehta_2023}, a theory-agnostic framework perturbs only the phase of gravitational-wave signals in standard tensor ($+, \times$) polarizations, without introducing scalar/vector modes or amplitude modifications. Phase deformations are parameterized and applied consistently through the inspiral, tapering off smoothly toward the merger--ringdown. The leading-order (–1 PN) deviation parameter, $\delta\phi_2$, captures potential dipole radiation. Posterior distributions from real and simulated signals are consistent with zero within 90\% credible bounds, indicating no significant deviation from GR. Similar analyses are performed in both GWTC-1~\cite{Abbott_2019BBH}, GWTC-2~\cite{Abbott_2021}, GWTC-3~\cite{GWTC3_2025} and GWTC-4~\cite{GWTC4_PTGR} where parameterized tests of GR introduce fractional
deviations $\delta \phi_n$ in the inspiral PN phase coefficients, one at a time, while keeping other coefficients at their GR values. Bayesian inference is used to constrain these deviations from LIGO--Virgo binary black hole signals. Across both catalogs,
\begin{figure*}[t]
    \centering
    \includegraphics[width=0.9\textwidth]{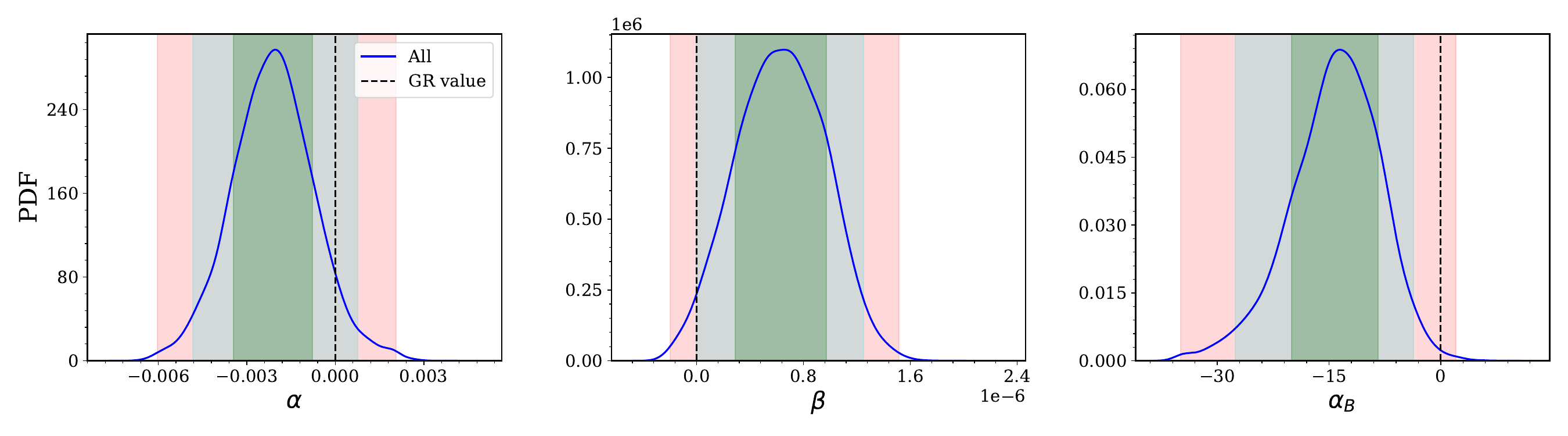}
    \caption{Distribution of the PPE parameters for the ``All" case for $\ell = |m|$ = 2 keeping a$_{\textrm T}$ =-2, a$_{\textrm S}$ = 0 and b = -7.}
    \label{fig4}
\end{figure*}
including the $-1$~PN term ($\delta \phi_{-2}$) sensitive to dipole radiation, the posteriors are consistent with zero with 90\% CI, showing no significant departures from GR. This can be compared to the phase modification in our analysis, where only the PPE parameter $\beta$ is considered alongside the 11 GR parameters, affecting only the phase of tensor modes without introducing scalar modes. The parameter $\beta$ is related to the $-1$~PN coefficient by $\beta = \frac{3}{128\eta}\, \delta \phi_{-2}$. As shown in Fig.~\ref{fig7}, $\beta$ and consequently $\delta \phi_{-2}$, exhibit a mild deviation from the GR value (i.e.\ $\beta=0$) which lies at the $\sim 96\%$ credible interval. This difference likely arises from the different methodologies considered, since we cut the waveform before the merger.

To compare the GR model (i.e.\ $\alpha=\alpha_B=\beta=0$) with the modified gravity PPE framework using the aforementioned dataset, we compute the difference in the $\chi^2$ statistic with respect to GR. The $\chi^2$ is defined as $\chi^2 = -2 \log L$, where $\log L$ denotes the maximum log-likelihood of the model. The model comparison is then performed via the likelihood ratio $\Delta \chi^2 = -2 (\log L_{\mathrm{GR}} - \log L_{\mathrm{PPE}})$. In addition, we evaluate model preference using the Akaike Information Criterion, defined as $\Delta \mathrm{AIC} = 2\,\Delta k + \Delta \chi^2$, where $\Delta k$ is the difference in the number of free parameters between the two models. For the GR model, we obtain a maximum log-likelihood $\log L_{\mathrm{GR}} = 559.49$, while for the PPE model we find $\log L_{\mathrm{PPE}} = 560.75$. The resulting likelihood difference $\Delta \log L = -1.26$ corresponds to $\Delta \chi^2 = 2.52$, indicating only a marginal improvement in the maximum likelihood for the PPE model, which is insufficient to compensate for its additional free parameters. Indeed, we also performed Bayesian model comparison, which yields a Bayes factor of $\Delta \log Z = 4.67 \pm 0.32$, providing \textit{strong evidence} in favor of General Relativity over the PPE extension according to Jeffrey's scale~\cite{Jeffreys1961} due to the penalization of extra parameters in the PPE model.

\begin{figure*}[t]
    \centering
    \includegraphics[width=0.9\textwidth]{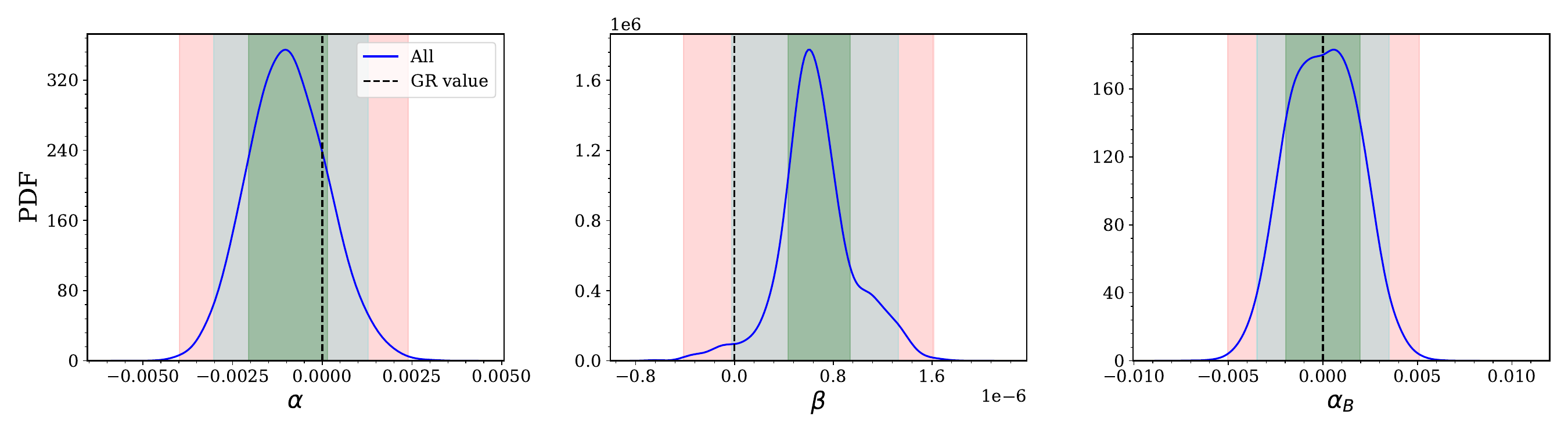}
    \caption{Distribution of the PPE parameters for ``All" case for $\ell = |m|$ = 1 keeping a$_{\textrm T}$ = a$_{\textrm S}$ = -2 and b = -7.}
    \label{fig5}
\end{figure*}
%%%%%%%%%%%%
\subsubsection{Frequency evolution: $a_T=-2$, $b=-7$, $a_S=0$}

Analogously to the previous section, we analyze the posterior distribution of the three PPE parameters $(\alpha,\beta,\alpha_B)$ when the breathing mode has a quadrupolar angular dependence. The difference is that now the frequency evolution of the scalar waveform has a power $a_S=0$ instead of $a_S=-2$. Notice that the tensor waveform has the same frequency evolution as in the previous case.

The qualitative properties of the posteriors are quite similar, and hence we only show the ``All" case in  Fig.~\ref{fig4}. Here also the breathing mode amplitude parameter $\alpha_B$ has double peaks for the case ``Sky+$d_L$+$\iota$'' which disappeared in the ``All" case as in the previous case. In the ``All" case we now obtain $\alpha=-2.11^{+2.87}_{-2.73} \times 10^{-3}$ at $95\%$ credible interval which improves $\sim 8\%$ by the inclusion of constraint on polarization angle. The phase parameter is constrained to $\beta=(6.33^{+6.17}_{-6.36}) \times 10^{-7}$ rad at $95\%$ and the EM data does not help improve much this constraint. 
The GR value $\alpha = 0$ lies near the median, while $\beta = 0$ is at a similar distance as in the $a_S = -2$ case; both parameters are consistent with GR within the 95\% credible interval.

For $\alpha_B = -13.79^{+10.14}_{-13.82}$ (95\% CI), the GR value $\alpha_B = 0$ lies at the \(\sim 99.2\%\) CI of the posterior, indicating a preference for scalar polarization, similar to the $a_S = -2$ case. Notice that the median value of \(\alpha_B\) seems to be very large compared to the previous case $a_S=-2$ but they are statistically the same and this value is consistent with the choice of \(a_S=0\) as can be seen from Tab.~\ref{tab1}. The inclusion of polarization angle improves the bounds on $\alpha_B$ by $\approx 35\%$. In addition, similar correlations are found among the parameters : \(r[{\beta-{\mathcal M}_c}]\approx -1\),  \(r[\alpha-d_L]\approx 0.6\),  \(r[\alpha-\alpha_B]\approx 0.3\) and \(r[\alpha_B-\psi]\approx 0.3\). 

%%%%%%%%%
\subsection{Dipole angular harmonic: $\ell = |m|=1$ }\label{sec:dipole}
In this section we analyze the scalar polarization in the dipole angular harmonic $\ell = |m| = 1$, keeping the tensor polarization in the quadrupole harmonic ($\ell = |m| = 2$) unchanged.

\begin{figure*}[t]
    \centering
    \includegraphics[width=0.9\textwidth]{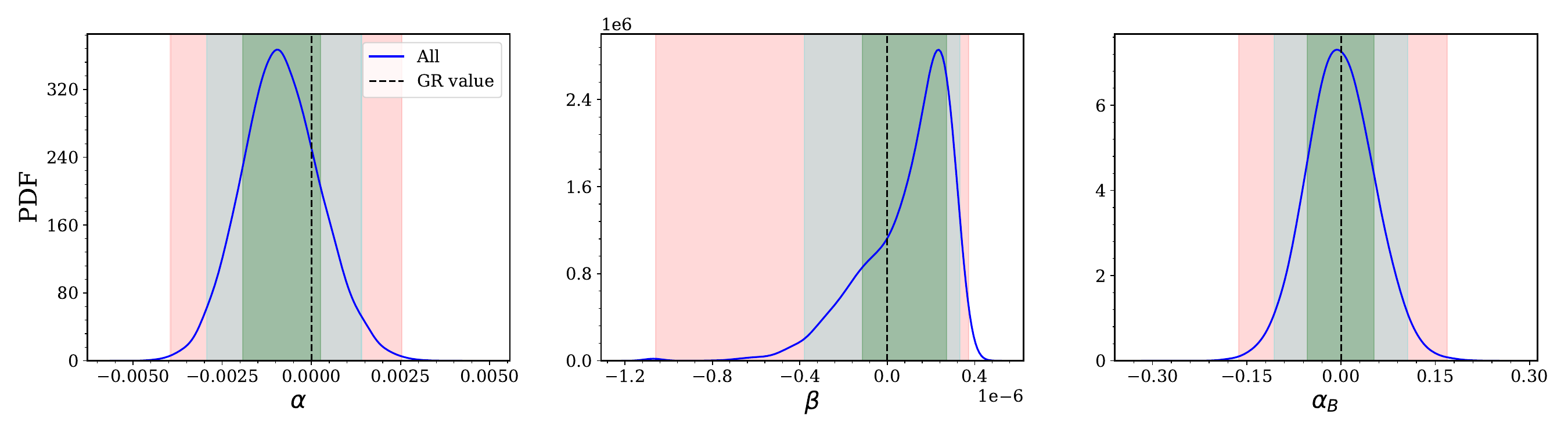}
    \caption{Distribution of the PPE parameters for ``All" case for $\ell = |m|$ = 1 keeping a$_{\textrm T}$ = -2, a$_{\textrm S}$ = 0 and b = -7.}
    \label{fig6}
\end{figure*}

%%%%%%%%%%%%%%%
\subsubsection{Frequency evolution: $a_T=-2$, $b=-7$, $a_S=-2$}

Fig.~\ref{fig5} shows the PPE parameter distributions for the ``All'' scenario. We obtain $\alpha = (-0.987^{+2.265}_{-2.047}) \times 10^{-3}$ and $\beta = 6.41^{+6.87}_{-6.61} \times 10^{-7}$ (both at 95\% CI), with $\beta$ comparable to the $\ell=|m|=2$ case. The scalar amplitude $\alpha_B = 0.01^{+3.50}_{-3.50} \times 10^{-3}$ (95\% CI) shows significant improvement over the $\ell=|m|=2$ case. All parameters are consistent with GR, with zeros well within 68\% CI. Compared to the $\ell=m=2$ case, the amplitude modification parameters have median values closer to their GR values with notably narrower distributions, except for the phase modification parameter $\beta$. This is attributed to the enhanced frequency evolution in both amplitude and phase of the scalar mode for $\ell=|m|=1$, the increased scalar amplitude from the $\sin\iota$ (vs.\ $\sin^2\iota$) factor, and the scalar mode cut-off frequency being half that of the $\ell=|m|=2$ case — collectively driving the non-GR PPE parameters toward GR consistency. Here we also found a strong correlation \(r[{\beta-{\mathcal M}_c}]\sim -1\), and  \(r[\alpha-d_L]\sim 0.8\). \(\alpha_B\) does not seem to be correlated with any of the parameters.

%%%%%%%%%%%%%%%%%%
\subsubsection{Frequency evolution: $a_T=-2$, $b=-7$, $a_S=0$}

We repeated the analysis with $a_S = 0$, incorporating all information via priors as in the $\ell = |m| = 2$ case. Fig.~\ref{fig6} shows the resulting PPE parameter distributions. The parameter $\alpha$ is constrained to $-0.878^{+2.284}_{-2.056} \times 10^{-3}$ (95\% CI), a 23\% improvement over the corresponding case of $\ell=|m|=2$, while $\beta = 1.505^{+1.819}_{-5.289} \times 10^{-7}$ (95\% CI) improves by 43\%. The scalar amplitude $\alpha_B = -0.002^{+0.108}_{-0.104}$ is well constrained within 68\% CI, showing significant improvement. All PPE parameters remain in strong agreement with GR (within 68\% CI), and the inter-parameter correlations are consistent with the $\ell = |m| = 1$ case. 

We also performed this analysis with uniform prior on polarization angle,$\psi$ to see its impact. It is found both amplitude modification parameters $\alpha$,$\alpha_B$ do not improved by the polarization angle prior and shows consistency with GR. However, the phase modification parameter $\beta$ improves by $9\%$ ($49\%$) when the polarization-angle prior is included in the ``All'' case for $a_T=a_S=-2,\, b=-7$ ($a_T=-2,\, a_S=0,\, b=-7$), while remaining consistent with GR.

%%%%%%%%%%%%%%%%%%%%%%%%%%%%%%%%%%
%CONCLUSIONS
%%%%%%%%%%%%%%%%%%%%%%%%%%%%%%%%%%
\section{Conclusion}\label{Conc}

General Relativity (GR) predicts only two tensor polarization modes (+ and $\times$), but in a generic metric theory a total of six polarization modes are possible. We performed a parameterized test of GR for the binary neutron star event GW170817, including the scalar breathing polarization along with the two tensor modes. We used the parameterized post-Einsteinian framework, which modifies the amplitude and phase of the tensor modes and includes the non-GR scalar polarization. The PPE parameterization for this case adds six extra non-GR parameters: three of them are fixed ($a_T$, $a_S$, $b$) which determine the frequency scaling of the modifications to GR and here we fixed to map specific modified gravity theories, while the rest of the parameters (\(\alpha, \beta, \alpha_B\)) determine the magnitude of the modifications and are kept free. These modifications are considered only for the inspiral part of the waveform, since there are no general predictions for modified gravity during the merger phase. We performed this analysis for two sets of angular harmonics $\ell = |m| = 1$ and  $\ell = |m| = 2$. For each case we fixed ($a_T$, $a_S$, $b$) to (-2, -7, -2) or to (-2, -7, 0).  

For the first time, we include a prior on the polarization angle for GW170817, coming from the gamma-ray burst afterglow. 
This results in significantly tighter constraints on non-GR PPE parameters, especially on the amplitude modifications of tensor and scalar polarizations. For instance, in the case $a_T = a_S = -2, b = -7$, the constraints improve by $60\%$ ($30\%$) for the scalar (tensor) mode amplitude for $\ell = |m| = 2$. In this case, the phase parameter $\beta$ remains consistent with GR, while the amplitude parameters $\alpha$ and $\alpha_B$ show mild tension from GR, i.e.\ their GR values $\alpha=0$ and $\alpha_B=0$ are at \(\sim 96.4\%\) CI and \(\sim 99.5\%\) CI respectively, when all the priors coming from the electromagnetic counterparts, including polarization angle, are applied.  While for \(a_T =-2, a_S=0, b=-7\) only the scalar mode amplitude parameter \(\alpha_B\) shows deviation from zero at the $99.2\%$ CI. Within the extended PPE model, the posterior for $\alpha_B$ favors non-zero values; however, Bayesian model comparison indicates that the simpler GR model remains preferred. On the contrary, for the angular harmonics $\ell = |m| = 1$ all the parameters are in good agreement with GR  independently of the choice of  \(a_T ,a_S\) and $b$. This arises from the more pronounced frequency evolution in both amplitude and phase of the scalar mode for $\ell = |m| = 1$, the larger scalar amplitude due to the $\sin\iota$ (as opposed to $\sin^2\iota$) dependence, and the fact that the scalar mode cut-off frequency is half that of the $\ell = |m| = 2$ case---together pushing the non-GR PPE parameters toward consistency with GR. Finally, comparing the two PPE waveform configurations considered in this work, the quadrupole-only configuration ($\ell = |m| = 2$) is decisively favored over the combined configuration ($\ell = |m| = 2$ and $\ell = |m| = 1$), with a log Bayes factor of $7.50 \pm 0.36$.

It remains unclear whether the mild preference for modified gravity is a real physical signal or just noise. Neutron stars can develop strong scalar effects, so binary neutron star mergers may show deviations from General Relativity that black hole mergers do not. Combining black hole events could therefore hide signals unique to neutron stars. The result from GW170817 may indicate such source-specific physics, which future observations of neutron star mergers could help confirm. Nevertheless, the current Bayesian evidence favors GR over the extended PPE model, implying that any apparent preference for non-zero modified gravity parameters should be regarded as tentative until  additional binary neutron star observations improve the statistical uncertainties.

\acknowledgments
We thank Thibeau Wouters, Kaze Wong and Thomas
C. K. Ng for useful discussions in setting up the \textsc{Jim} runs. This project was funded by ESO-Chile Comité Mixto, and Fondecyt Iniciación 11250105 grant.

\section{Data Availability}
The code and results supporting the findings of this study publicly available \cite{jim_v1}. 
The posterior samples of this work are publicly available on Zenodo~\cite{Imam2026_ns}.
\bibliographystyle{apsrev4-1}
% \bibliography{Refs.bib}

% \pagebreak
% \newpage

%%%%%%%%%%%%%%%%%%%%%%%%%%%%%%%%%%
%APPENDIX
%%%%%%%%%%%%%%%%%%%%%%%%%%%%%%%%%%
\appendix

\begin{figure*}
  %  \centering
\includegraphics[width=0.3\linewidth]{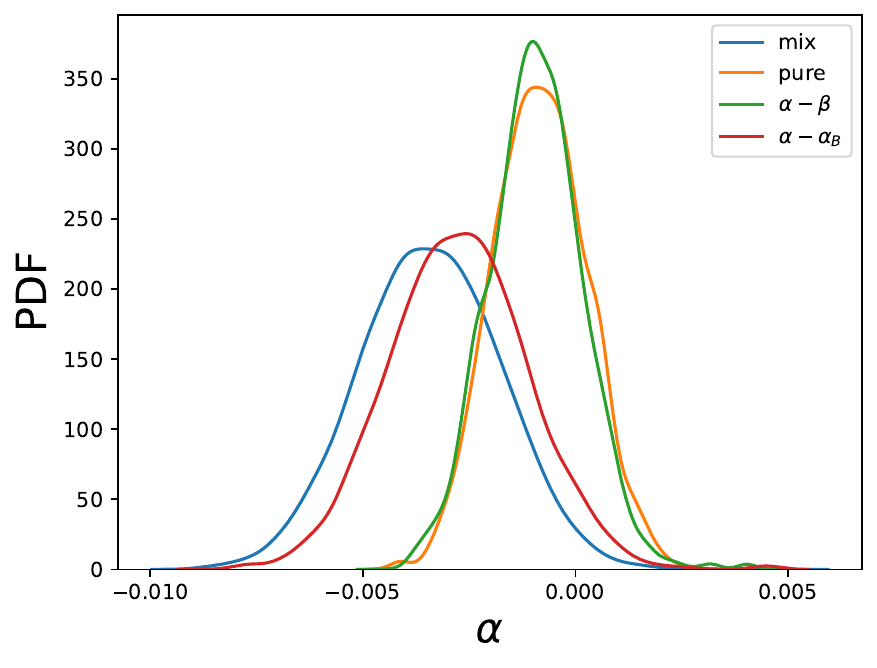}
\includegraphics[width=0.3\linewidth]{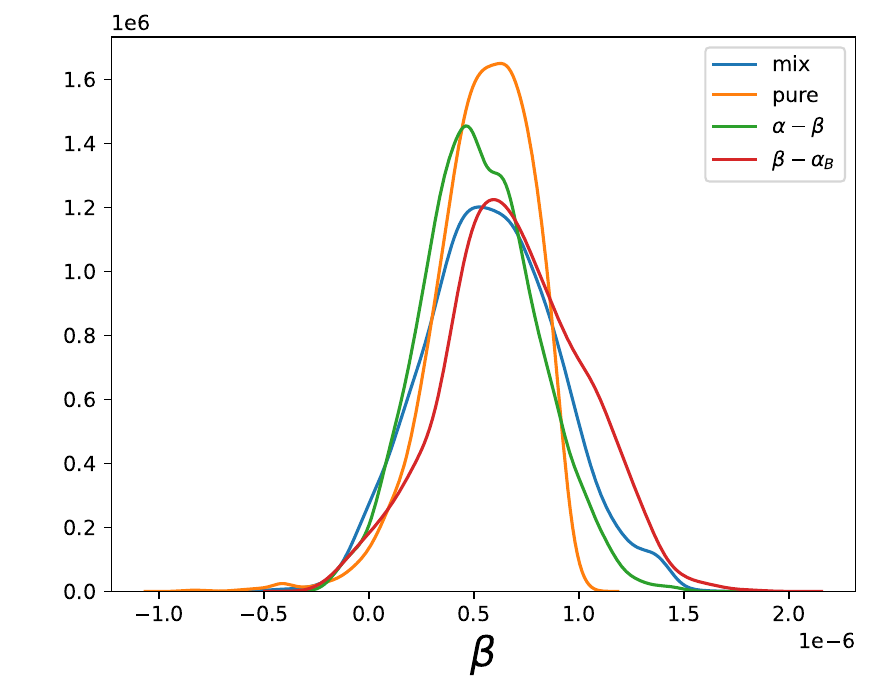}
\includegraphics[width=0.3\linewidth]{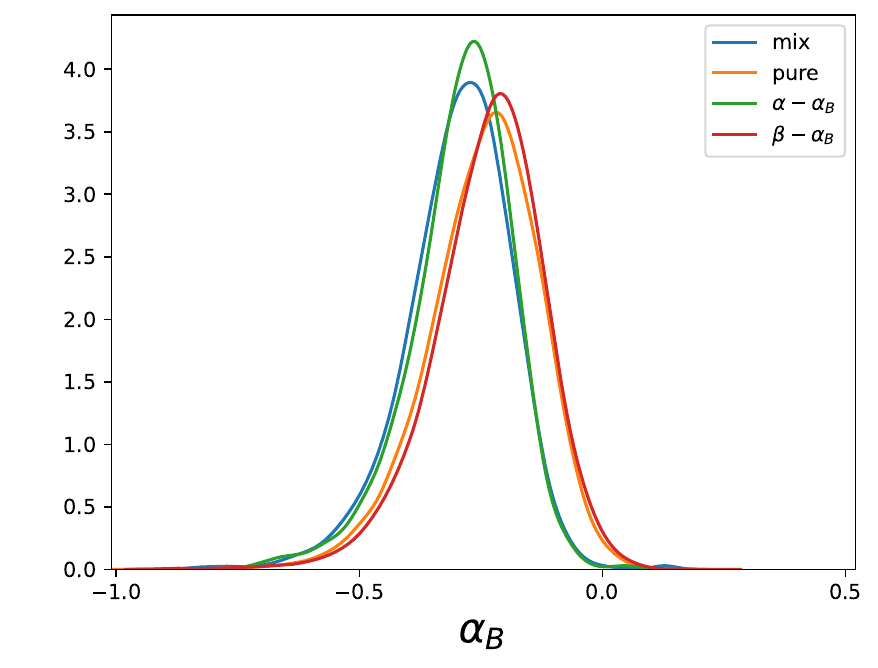}
\caption{Distribution of the parameter \(\alpha_B\) with all the information for (i) mix : when all the three parameters (ii) pure : only \(\alpha_B\) (iii) \(\alpha-\alpha_B\) and (iv)\(\beta-\alpha_B\) presents in the waveform for $\ell = |m|$ = 2 keeping a$_{\textrm T}$ = a$_{\textrm S}$ = -2 and b = -7.}
    \label{fig7}
\end{figure*}

\begin{figure*}[t!]
\includegraphics[width=0.9\linewidth]{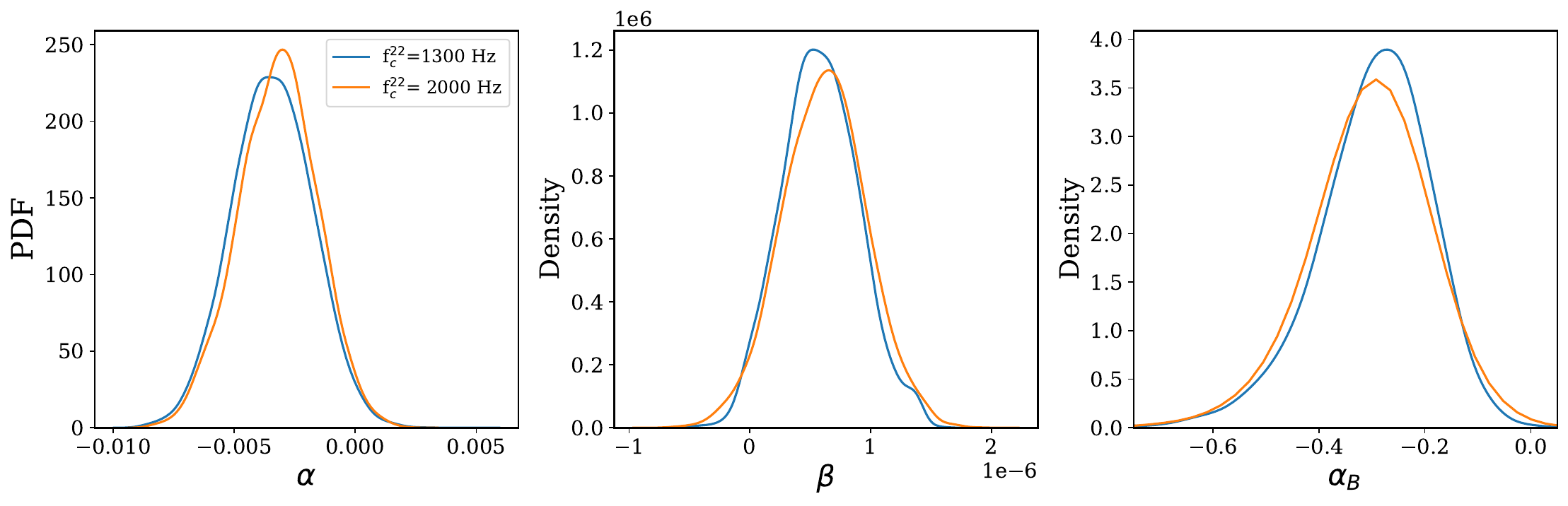}
\caption{Dependency of the PPE parameters on cut off frequency for the ``All" case for $\ell = |m|$ = 2 keeping a$_{\textrm T}$ = a$_{\textrm S}$ = -2 and b = -7.}
    \label{fig8}
\end{figure*}
\section{Parameter Degeneracy}\label{Appendix_A}

In Fig.~\ref{fig2}, there are double peaks in the breathing-mode amplitude parameter, $\alpha_B$ when the polarization angle priors are uniform. This is related to the $\pi/2$ symmetry of the polarization angle which can be analytically explained:

\bea
h = F_+(\psi) \tilde{h}_+ + F_\times(\psi) \tilde{h}_\times + F_b \tilde{h}_b
\eea
For \(\ell=m=\)2 :
\begin{align}
\tilde{h}_{+/\times}^{(2,2)}(f)&=  \tilde{h}_{\rm GR,+/\times}^{(2,2)}(f) (1 + \alpha u_2^{a_T})e^{i\beta u_2^{b}},\\
& = \tilde{h}^{(2,2)}_0(1 + \alpha u_2^{a_T})e^{-i\psi_T}
\end{align}
with the phase \(\psi_T =\psi_{GR} -\beta u^b_2\) and \(\tilde{h}^{(2,2)}_0\)is the amplitude of the tensor modes.
Similarly,
\begin{align}
\tilde{h}^{(2,2)}_b &=\tilde{h}^{(2,2)}_{0b}\alpha_B u_2^{a_S}e^{-i\psi_T}
\end{align}
Now, considering \[k_1 = F_+\tilde{h}^{(2,2)}_0(1+\alpha u^{a_T}_2)\] , 
\[k_2 = F_\times\tilde{h}^{(2,2)}_0(1+\alpha u^{a_T}_2)\] 
and \[k_3 = F_b\tilde{h}^{(2,2)}_{0b}\alpha_B u^{a_S}_2\]

\begin{align}
h&= \left(k_1(1+\xi^2) + 2k_2 i\xi + k_3(1-\xi^2)\right)exp(-i\psi_T),\\
& = (A + iB)exp(-i\psi_T),\\
& = r exp(-i\psi_T+i\Delta\theta)
\end{align}
Here \[A=k_1(1+\xi^2) + k_3(1-\xi^2) \] and \[B = 2k_2\xi \]
So,
\begin{align}
 r &=\sqrt{A^2+B^2},\\
   &=\sqrt{k^2_1(1+\xi^2)^2 + k^2_3(1-\xi^2)^2 + 2k_1k_3(1-\xi^4) + 4k^2_2\xi^2}
\end{align}
\begin{align}
\Delta\theta &= arctan
\left(\frac{B}{A}\right),\\
&=arctan\left(\frac{2k_2\xi}{k_1(1+\xi^2) + k_3(1-\xi^2)}\right)
\end{align} 
so when $\psi\rightarrow\psi + \pi/2$ the antenna pattern functions $F_+, F_\times$, i.e., $k_1, k_2$ change sign, the phase remains unchanged if $k_3$ also flips sign. Now, as $F_b$ does not include $\psi$, a sign change can only happen if $\alpha_B$ flips sign. So the double peaks appear due to the $\pi/2$ symmetry of the polarization angle, and when the polarization angle is fixed, the double peaks in $\alpha_B$ disappear.

\section{Pure vs mixed polarization}\label{Appendix_B}
We studied deviations from GR by incorporating all three non-GR parameters into the waveform i.e., considering mixed polarizations as well as including only a single non-GR parameter to represent pure polarization cases. From Fig.~\ref{fig7}, it is evident that the deviations are more significant when mixed polarization modes are considered, compared to the cases with pure polarization modes, specifically for the parameter \(\alpha\). To identify the underlying cause, we remove one parameter and analyze using two parameters at a time. From Fig.~\ref{fig7} it is clear that whenever \(\alpha\) and \(\alpha_B\) are included, their distributions deviate more from GR predictions. This behavior arises from the intrinsic correlation between \(\alpha\) and \(\alpha_B\) along with the constraints on \(\psi\). Similar deviations are seen in $\alpha_B$ when \(\alpha\) and \(\alpha_B\) are included together. But the distribution of $\beta$ remains the same for all cases.

\section{Cut off frequency dependency}\label{Appendix_C}

The modifications are introduced only in the inspiral part of the waveform. So we terminated the waveform above a cutoff frequency \( f^{(2,2)}_{\text{c}} = 0.018/M_z \approx 1300\) Hz, 
where \( M_z =2.78 M_\odot \) is the redshifted total mass of the binary. To test the dependence of the results on the cut-off frequency, 
we manually increased \( f^{(2,2)}_{\text{c}} \) to 2000~Hz. As shown in Fig.~\ref{fig8},
the results remain unchanged with respect to the cutoff frequency for 
\( a_T = a_S = -2 \), owing to the presence of the terms \( u^{a_T} \) and \( u^{a_S} \) in the waveform amplitudes.

\begin{figure*}[t!]
    \centering
    \includegraphics[width=0.85\textwidth]{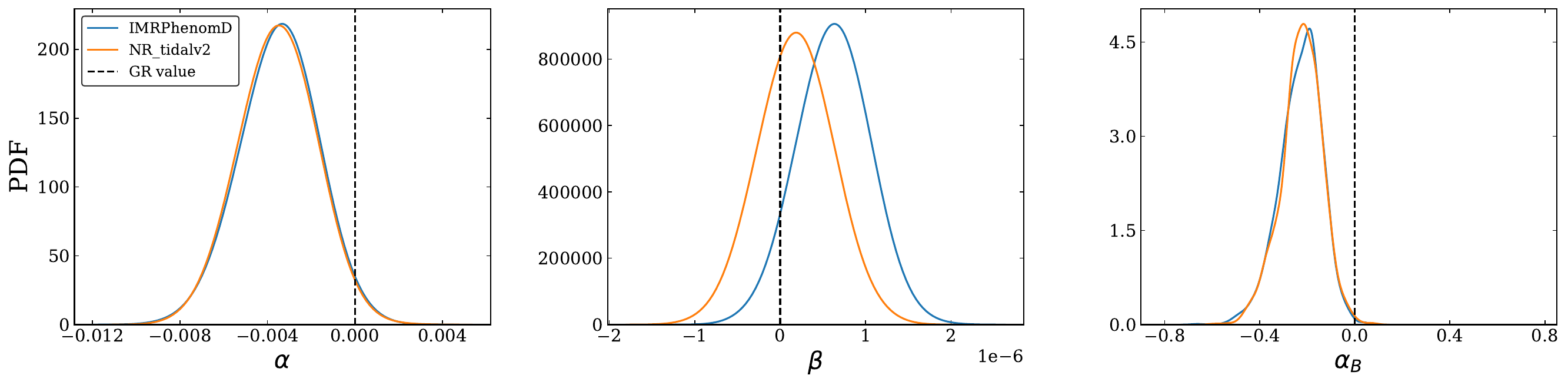}
    \caption{1D posterior distributions of PPE parameters ($\alpha$, $\beta$, $\alpha_B$) comparing IMRPhenomD (blue) and NRTidalv2 (orange) waveform models. Black dashed lines indicate GR values. }
    \label{fig9}
\end{figure*}

\begin{figure*}[t!]
    \centering
    \includegraphics[width=0.82\textwidth]{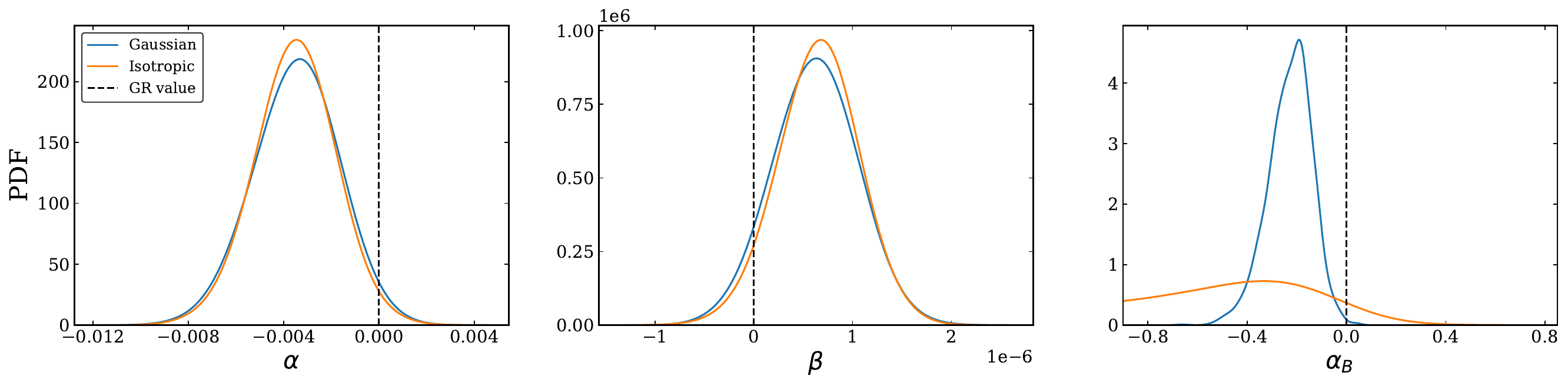}
    \caption{Prior sensitivity analysis for orbital inclination $\iota$. Comparison of isotropic and Gaussian prior ( $\mu=2.85$, $\sigma=0.03$) posteriors.}
    \label{fig10}
\end{figure*}

\section{IMRPhenomD vs Tidal waveform}\label{Appendix_D}
We have performed the complete analysis using both \texttt{IMRPhenomD} (point-particle) and \texttt{IMRPhenomD\_NRTidalv2} (tidal) waveforms to assess the impact of tidal effects on our PPE parameter constraints. Fig.~\ref{fig9} presents the posterior distributions of the three PPE parameters ($\alpha$, $\beta$, $\alpha_B$) obtained with each waveform model. The posteriors of the PPE parameters  remain consistent between the two waveforms, with negligible shifts in $\alpha$ and $\alpha_B$. Although the tidal waveform modifies the $\beta$ posterior distribution, resulting in a Jensen--Shannon divergence of $D_{\rm JS}=0.15$ between the two analyses. This indicates a moderate change in the inferred posterior while the two distributions remain broadly consistent. Most importantly, the mild preference for scalar breathing modes ($\sim 2$--$3\sigma$ deviation from zero) is robust to the choice of waveform model, confirming that our scalar-mode constraint is not an artifact of unmodeled tidal physics.

\section{Inclination angle prior dependency}\label{Appendix_E}
We compare two different prior choices on the binary inclination angle ($\iota$): the standard isotropic sine-based prior and a Gaussian prior informed by electromagnetic observations ($\text{Gaussian}(\mu = 2.85, \sigma = 0.03)$ rad, derived from VLBI and gamma-ray burst light curve data). Fig.~\ref{fig10} presents the posterior distributions of the PPE parameters ($\alpha$, $\beta$, $\alpha_B$) obtained under each prior specification. The posteriors on the PPE parameters \(\alpha, \beta\) show consistency between the two priors. While for the Gaussian prior prior \(\alpha_B\) posterior is very narrow compared to the isotropic prior with \(D_{JS}=0.36\), the combination \(\alpha_B \sin^2(\iota)\) which appears in the breathing mode amplitude has \(D_{JS}=0\).

\section{Harmonic Mode Comparison: Quadrupole vs. Dipole-Quadrupole Configuration}\label{Appendix_F}
\begin{figure*}
    \centering
    \includegraphics[width=0.9\textwidth]{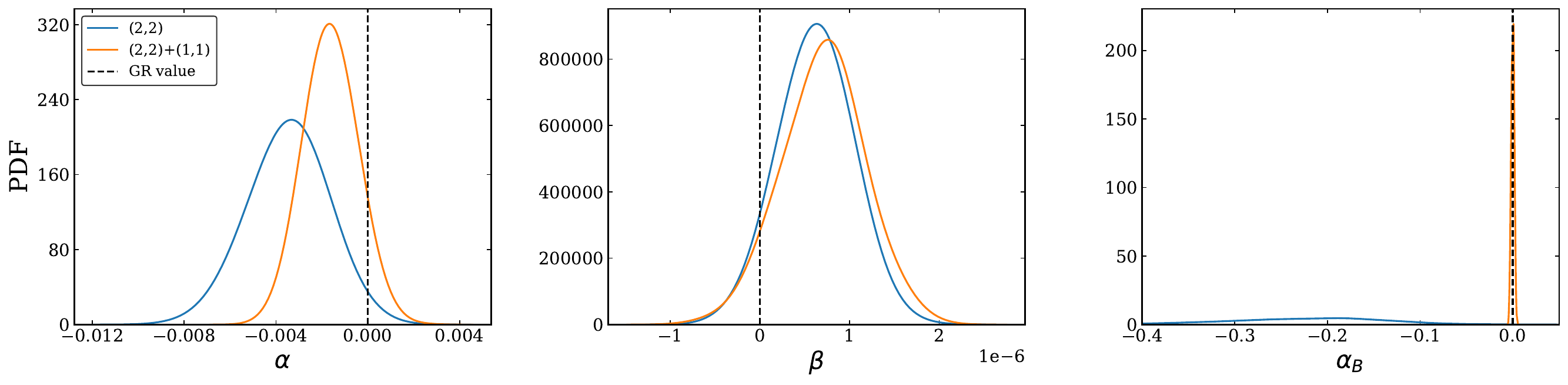}
    \caption{Posterior distributions of PPE parameters comparing breathing 
    mode parametrizations: (2,2) mode alone (blue) versus (2,2)+(1,1) combined 
    modes (orange). Although the (2,2)+(1,1) combination yields a tighter 
    constraint on $\alpha_B$ (closer to GR), Bayes factor analysis 
    ($\ln(\mathcal{B}) = 7.504 \pm 0.363$) decisively favors the simpler 
    (2,2) mode, indicating that the precision gain from the dipole mode does 
    not justify the added model complexity for GW170817.}
    \label{fig11}
\end{figure*}

We compare the two harmonic configurations of the scalar breathing mode: the quadrupole-only case ($\ell = |m| = 2$) and the combined dipole-quadrupole case ($\ell = |m| = 2$ and $\ell = |m| = 1$). Fig.~\ref{fig11}  presents the posterior distributions of the three PPE parameters ($\alpha$, $\beta$, $\alpha_B$) for both configurations. The Bayesian model comparison between the two scenarios yields a log Bayes factor, $\Delta \log Z = 7.50 \pm 0.36$, with decisive evidence favoring the quadrupole-only configuration, with $\log Z_{(2,2)} = 526.49 \pm 0.24$ and $\log Z_{(2,2)+(1,1)} = 518.99 \pm 0.27$. According to Jeffrey's scale, this Bayes factor ($\Delta \log Z > 5$) constitutes decisive evidence for the simpler quadrupole-only model. Notably, while the combined dipole-quadrupole configuration yields a tighter constraint on the amplitude modification parameters \(\alpha, \alpha_B\) with \(D_{JS}=0.21, 0.69\), respectively, the data strongly prefer the less complex model. This result is confirmed with both \texttt{IMRPhenomD} and \texttt{IMRPhenomD\_NRTidalv2} waveforms, demonstrating the robustness of this finding.

\section{All Parameters}\label{Appendix_G}
\begin{figure*}
    \centering
    \includegraphics[width=\linewidth]{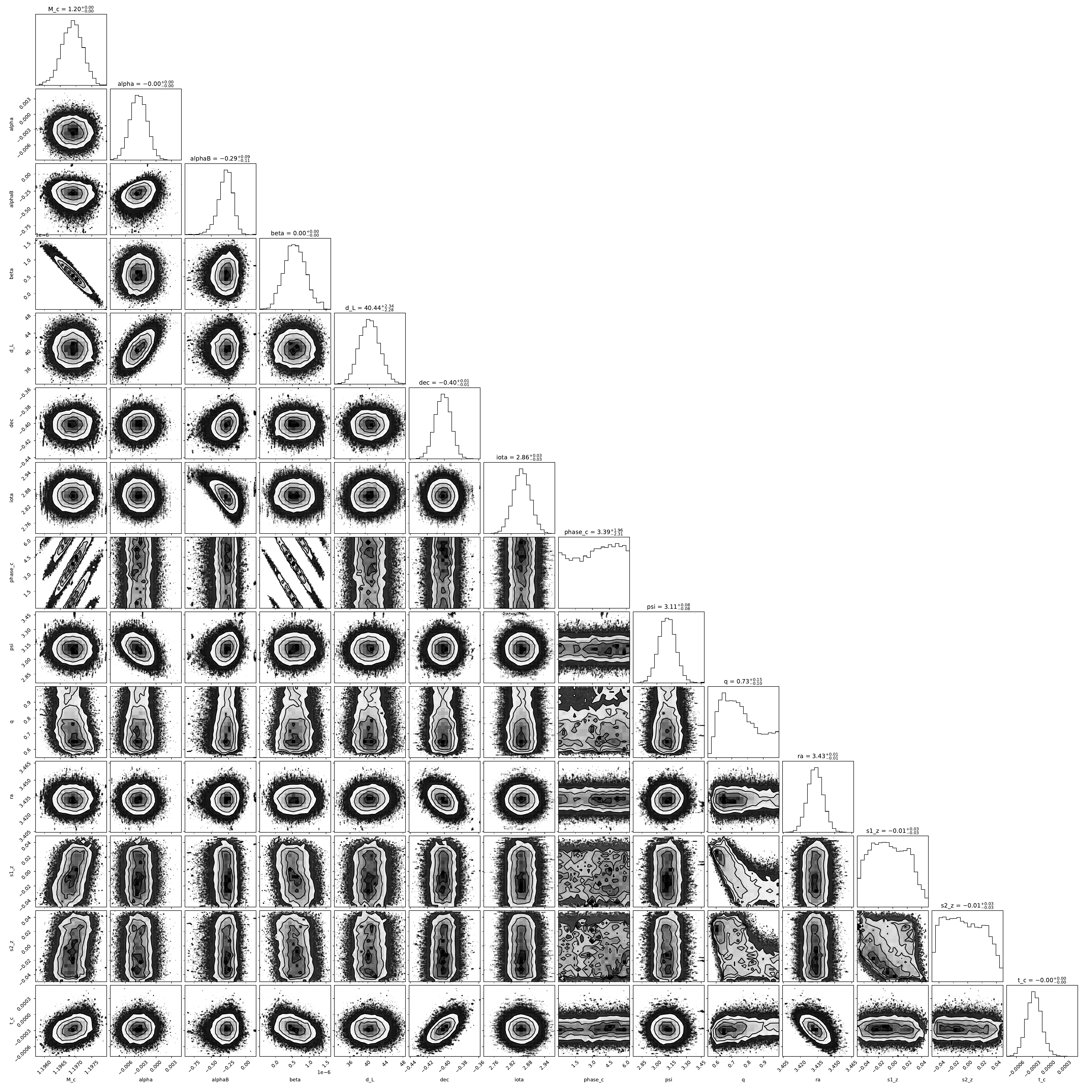}
    \caption{Corner plot for the ``All" case for $\ell = |m|$ = 2 keeping $a_{\textrm T} = a_{\textrm S} = -2$ and $b = -7$.}
    \label{fig12}
\end{figure*}

In Fig~\ref{fig12} the corner plot of the 3 non-GR parameters along the 11 GR parameters are shown for the ``All" case for the angular harmonics $\ell=m=2$ and keeping \(a_T= a_S\)=-2 and b =-7.

\end{document}